\documentclass{aastex63}
\usepackage{savesym}
\savesymbol{tablenum}
\restoresymbol{SIX}{tablenum}

\newcommand \ang[1]{#1$^\circ$}

\shortauthors{The HAWC Collaboration}

\graphicspath{{./}}

\begin{document}

\title{3HWC: The Third HAWC Catalog of Very-High-Energy Gamma-ray Sources}



\correspondingauthor{Henrike Fleischhack}
\email{hfleisch@mtu.edu} 

\correspondingauthor{Mehr Un Nisa}
\email{nisamehr@msu.edu} 

\correspondingauthor{Alison Peisker}
\email{peiskera@msu.edu} 


\author[0000-0003-0197-5646]{A.~Albert}
\affiliation{Physics Division, Los Alamos National Laboratory, Los Alamos, NM, USA }
\author[0000-0001-8749-1647]{R.~Alfaro}
st\affiliation{Instituto de F\'{i}sica, Universidad Nacional Autónoma de México, Ciudad de Mexico, Mexico }
\author{C.~Alvarez}
\affiliation{Universidad Autónoma de Chiapas, Tuxtla Gutiérrez, Chiapas, México}
\author{J.D.~Álvarez}
\affiliation{Universidad Michoacana de San Nicolás de Hidalgo, Morelia, Mexico }
\author{J.R.~Angeles Camacho}
\affiliation{Instituto de F\'{i}sica, Universidad Nacional Autónoma de México, Ciudad de Mexico, Mexico }
\author{J.C.~Arteaga-Velázquez}
\affiliation{Universidad Michoacana de San Nicolás de Hidalgo, Morelia, Mexico }
\author{K.P.~Arunbabu}
\affiliation{Instituto de Geof\'{i}sica, Universidad Nacional Autónoma de México, Ciudad de Mexico, Mexico }
\author{D.~Avila Rojas}
\affiliation{Instituto de F\'{i}sica, Universidad Nacional Autónoma de México, Ciudad de Mexico, Mexico }
\author[0000-0002-2084-5049]{H.A.~Ayala Solares}
\affiliation{Department of Physics, Pennsylvania State University, University Park, PA, USA }
\author[0000-0003-0477-1614]{V.~Baghmanyan}
\affiliation{Institute of Nuclear Physics Polish Academy of Sciences, PL-31342 IFJ-PAN, Krakow, Poland }
\author[0000-0003-3207-105X]{E.~Belmont-Moreno}
\affiliation{Instituto de F\'{i}sica, Universidad Nacional Autónoma de México, Ciudad de Mexico, Mexico }
\author[0000-0001-5537-4710]{S.Y.~BenZvi}
\affiliation{Department of Physics \& Astronomy, University of Rochester, Rochester, NY , USA }
% ...

\collaboration{106}{(HAWC Collaboration)}

\begin{abstract}

We present a new catalog of TeV gamma-ray sources using 1523 days of data from the High Altitude Water Cherenkov (HAWC) observatory. The catalog represents the most sensitive survey of the Northern gamma-ray sky at energies above several TeV, with three times the exposure compared to the previous HAWC catalog, 2HWC. We report 65 sources detected at $\geq$ 5 sigma significance, along with the positions and spectral fits for each source. The catalog contains eight sources that have no counterpart in the 2HWC catalog, but are within \ang{1} of previously detected TeV emitters, and twenty  sources that are more than \ang{1} away from any previously detected TeV source. Of these twenty new sources, fourteen have a potential counterpart in the fourth \textit{Fermi} Large Area Telescope catalog of gamma-ray sources. We also explore potential associations of 3HWC sources with pulsars in the ATNF pulsar catalog and supernova remnants in the Galactic supernova remnant catalog.


\end{abstract}

\keywords{High-energy astrophysics, Gamma-ray astronomy --- catalogs --- surveys}


\section{Introduction} \label{sec:intro}


High-sensitivity, unbiased surveys of the gamma-ray sky are important to finding new astrophysical objects -- both to understand their bulk properties, and to constrain new physics beyond the standard model. Discovering the sources of cosmic rays and determining the underlying acceleration mechanisms requires precise measurements of gamma-ray spectra of objects above several tens of TeV. In addition, the indirect search for dark matter particles in the GeV--TeV regime also hinges on detecting a steady flux of photons from several galactic and extra-galactic targets of interest. The current generation of ground-based gamma-ray telescopes, in particular Imaging Atmospheric Cherenkov Telescopes \citep{Holder:2008ux,2014APh....54...67B,2017APh....94...29A,2013JInst...8P6008A}, are capable of resolving gamma-ray sources to $\leq 0.1^\circ$ precision. The highly complementary survey instruments, such as Tibet-AS$\gamma$ \citep{PhysRevLett.123.051101}, LHAASO \citep{Zhen:2014zpa}, ARGO-YBJ \citep{Bartoli:2013qxm}, and HAWC, have further extended the reach of TeV astronomy with their high up-time and unprecedented sensitivity to spatially extended sources.      

The High Altitude Water Cherenkov (HAWC) observatory has been continuously monitoring the Northern sky in TeV cosmic rays and gamma rays since commencing full operations in 2015, and has achieved a sensitivity down to a few percent of the Crab flux in five years. This work presents the results of an all-sky time-integrated search for point-like and extended sources using 1523 days of HAWC data. As a follow-up to the 2HWC catalog of TeV gamma-ray sources \citep{2HWC}, we introduce an updated catalog using more data and improved analysis methods.

This paper is structured as follows. Section \ref{sec:hawc} provides a brief description of the HAWC detector, as well as the data and the analysis method we use in the construction of the catalog. Section \ref{sec:results} presents the results of the catalog search and provides preliminary spatial and spectral information on all sources. A broad discussion of the results is also presented. Section \ref{sec:discuss} discusses the systematic uncertainties and methodological limitations of this work. Section \ref{sec:conclude} concludes the paper.

\section{Instrument and Data Analysis}\label{sec:hawc}
\subsection{The HAWC Gamma-ray Observatory} 

HAWC consists of 300 water tanks, each filled with $\sim$ 200,000 liters of purified water and instrumented with four photo-multiplier tubes (PMTs). Very-high-energy primary particles interact with Earth's atmosphere and create an extensive air shower of secondary particles. Charged particles produced in an air shower emit Cherenkov radiation as they pass through the water in HAWC's tanks, which in turn produces photo-electrons in the PMTs.

The HAWC observatory is sensitive to gamma rays in an energy range from hundreds of GeV to hundreds of TeV, with the exact energy threshold depending on the declination and energy spectrum of each source. Due to its location at a latitude of 19$^\circ$ N and its wide field-of-view, HAWC can observe about two thirds of the gamma-ray sky (from about \ang{-26} to +\ang{64} in declination) every day, with an instantaneous field-of-view of $>$2.0 sr. HAWC's angular resolution (68\% containment radius for photons) ranges between \ang{0.1} and \ang{1.0} depending on the energy and zenith angle of the signal. More details about the HAWC detector can be found in \cite{2HWC,oldCrab}.

\subsection{Data Selection and Reduction}

The temporal and spatial distribution of charge deposited in HAWC's PMTs is used to reconstruct the properties of the primary particle producing the air shower. The difference in timing between the signals recorded in different tanks allows us to reconstruct the direction of the primary particle. The spatial distribution and magnitude of the charges can be used to reconstruct the primary energy and to separate gamma-ray induced showers from cosmic-ray induced ones. 

We distribute the reconstructed events into nine analysis bins according to the fraction of the operating PMTs that recorded a signal for a given event. The fraction of PMTs hit is correlated with the primary energy, allowing us to extract the energy spectrum of gamma-ray sources. We apply gamma-hadron separation cuts to reduce the background of cosmic-ray induced showers. A detailed description of air shower reconstructions and quality cuts applied to the data can be found in \cite{2HWC,oldCrab}.

We further bin the gamma-ray candidate events according to the direction of the primary particle in celestial coordinates (Right Ascension and Declination, J2000 epoch). We use the HEALPix binning scheme \citep{healpix}, with an NSIDE parameter of 1024. 

The dominant background is given by hadronic showers that pass the gamma-hadron cuts. For each pixel, we estimate the expected number of remaining background events after cuts using the method of Direct Integration as described in \cite{DI_milagro}. 

\subsection{Construction of the Catalog} \label{sec:analysis}
The 3HWC catalog is based on data collected by the HAWC observatory between November 2014 and June 2019, corresponding to a livetime of 1523 days -- about three times the livetime of the 2HWC catalog data set. For the most part, the construction of the catalog follows the same method as the previous 2HWC catalog, which is described in \cite{2HWC}. We summarize the algorithm below, with particular focus on the differences from the previous catalog search.

\subsubsection{Source Search}

We perform a blind search for sources across HAWC's field-of-view using the likelihood framework discussed in \cite{liff}. The likelihood calculation assumes that the number of counts in each bin/pixel is distributed according to a Poisson distribution, with the mean given by the estimated background counts plus (if applicable) the predicted number of gamma-ray counts from the convolution of the source model with the detector response.

For each HEALPix pixel, we calculate a likelihood ratio $\lambda = \hat{\mathcal{L}}_{s+b}/\mathcal{L}_b$, comparing the likelihood, $\hat{\mathcal{L}}_{s+b}$, of the best-fit model with a gamma-ray source centered on that pixel to that of a background-only model, $\mathcal{L}_{b}$. We define a test statistic, $TS=2\,\log\left( \Lambda \right)$. Assuming that the null hypothesis is true, the $TS$ is distributed according to a $\chi^2$ distribution with one degree of freedom \citep{wilks1938}, which can be approximated by a gaussian distribution. Then, $\pm \sqrt{TS}$ corresponds to the (``pre-trials'') significance. The negative sign is used for pixels in which the best-fit flux normalization is negative. 

\begin{figure}[b]
\includegraphics[width=\textwidth, trim=0cm 0cm 0cm 3cm, clip=true]{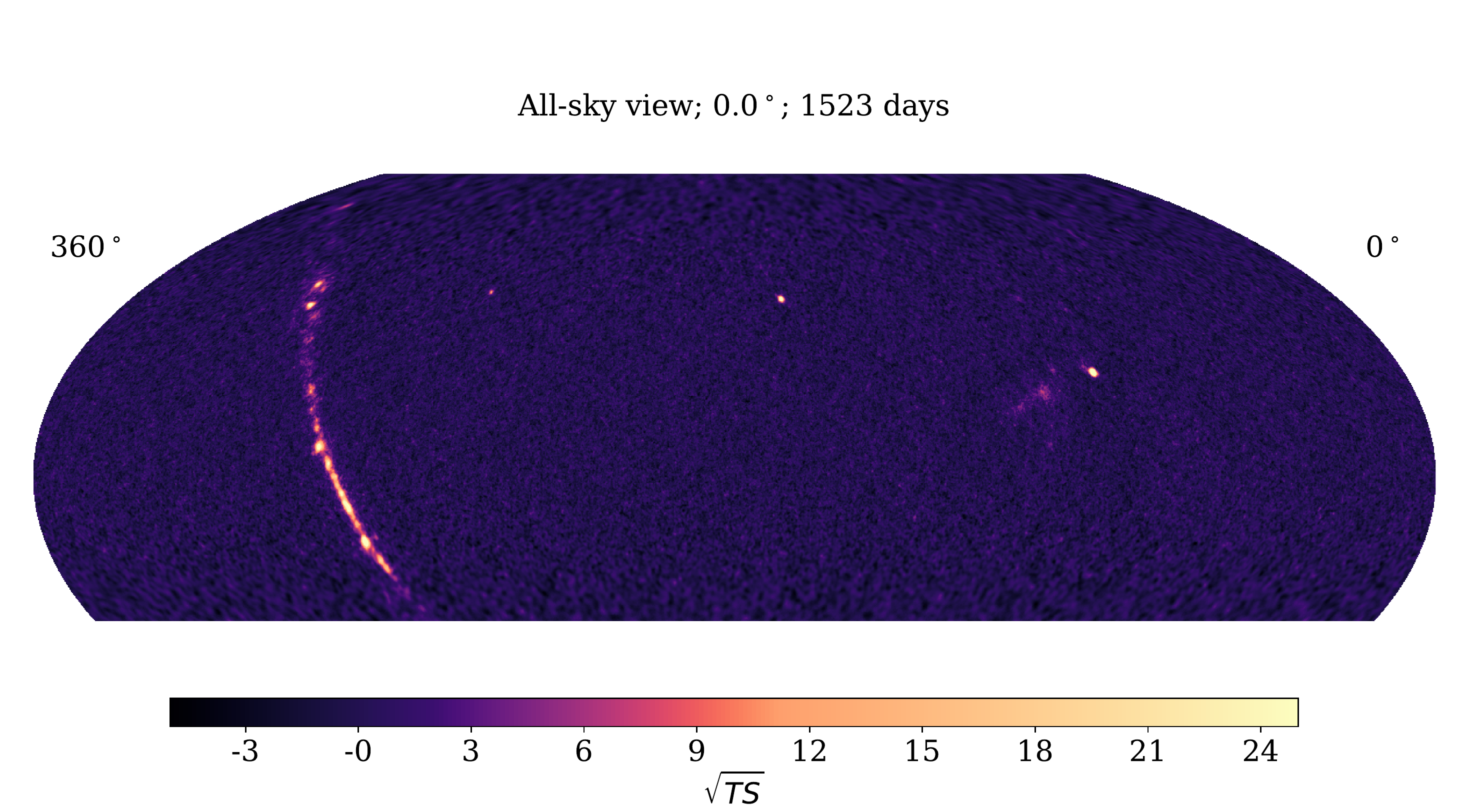}
\caption{All-sky significance map in celestial coordinates, assuming a point-source hypothesis. The bright band on the left is part of the Galactic plane (c.f. Figures \ref{fig:plane1}-\ref{fig:plane4}), and the bright region on the right is the Galactic anti-center region containing the Crab Nebula and the Geminga halo (c.f. Figure \ref{fig:plane5}). The two off-plane hotspots are the two TeV-bright blazars Mrk 421 (right) and Mrk 501 (left).}
\label{fig:allsky}

\end{figure}

The signal hypothesis considers a fixed source morphology and an $E^{-2.5}$ power-law energy spectrum. The only free parameter of the likelihood fit is the flux normalization. We repeat source searches for four different hypothetical morphologies: point sources, and extended disk-like sources with radii of \ang{0.5}, \ang{1.0}, and \ang{2.0}. This procedure is very similar to what was used in the previous 2HWC catalog, with the only change being the spectral index hypothesis (the prior catalog used a spectral index of -2.7 for point sources, -2.0 for extended sources). The resulting all-sky significance map for a point-source assumption can be seen in Figure \ref{fig:allsky}. 


For each significance map, we compile a list of candidate sources comprising the local maxima with $\sqrt{TS} > 5$. Due to Poisson fluctuations in the number of detected gamma rays, a single source may produce multiple local maxima. To avoid double-counting such sources, candidate sources are promoted to sources if they pass the  following ``TS valley'' criterion: the significance profile connecting the source candidate with any other source within \ang{5} of the source candidate in question has to ``dip'' by $\Delta TS>2$ to promote a source candidate to a ``primary'' source status. A ``secondary'' source is defined with a relaxed criterion, such that $1 < \Delta TS < 2$. We mark secondary sources with a dagger (\dag) in Table \ref{tab:sources}. The $\Delta TS$ criterion used for 3HWC is a little stricter than the one used in the 2HWC catalog, in which a source only had to pass the $\Delta TS$ with its closest neighboring source.

We then combine the four source lists (for the four different assumptions of the source morphology) to yield the 3HWC catalog. We include all sources found in the point source search in 3HWC. We only include sources found in the extended source searches if they are more than \ang{1} away from any point source or smaller extended source already in the catalog. 
Table \ref{tab:sources} shows the resulting list of sources comprising the 3HWC catalog. For each source, we also show the closest known TeV source listed in the TeVCat\footnote{\url{http://tevcat.uchicago.edu/}}  \citep{2008ICRC....3.1341W} if it is within \ang{1} of the HAWC source. 

\subsubsection{Spectral Fits}
After identifying the primary and secondary sources, we perform likelihood fits to obtain each source's energy spectrum. We assume a simple power-law spectrum for each source, fitting for the spectral index and flux normalization as done for the 2HWC catalog \citep{2HWC}. There are two changes to the spectral fitting procedure used in this work compared to 2HWC. First, for 3HWC, we only report the spectral fit for the same morphology for which the source was first found in the source-search stage. Second, we dynamically treat the instrument response during the fit. 2HWC fits relied on a method where the angular resolution was pre-calculated (and fit with a double-Gaussian function in each analysis bin) before the spectral fit, for a fixed spectral assumption. 3HWC spectral fits use a new method in which we recalculate the angular resolution for each tested spectral assumption during the fit. Additionally, we do not assume that the angular resolution follows a specific analytical shape. This allows for a more complete characterization of HAWC's PSF and of the systematic uncertainties on the fit parameters. See \cite{israelThesis} for more details on the new spectral fit method. Table \ref{tab:fluxes} shows the results of the spectral fits.


\section{Results}\label{sec:results}

\subsection{3HWC Sources}
The 3HWC catalog contains 65 sources, 17 of which are considered secondary sources (not well separated from neighboring sources according to the $\Delta TS$ criterion). The source positions can be found in Table \ref{tab:sources} and the results of the spectral fits, as well as the energy range from which we expect 75\% of the observed significance, can be found in Table \ref{tab:fluxes}. Twenty-eight of these sources do not lie within \ang{1} of any 2HWC source. We discuss some of these sources in more detail in Section \ref{sec:new_sources}.

\begin{figure}[tbp]
\plotone{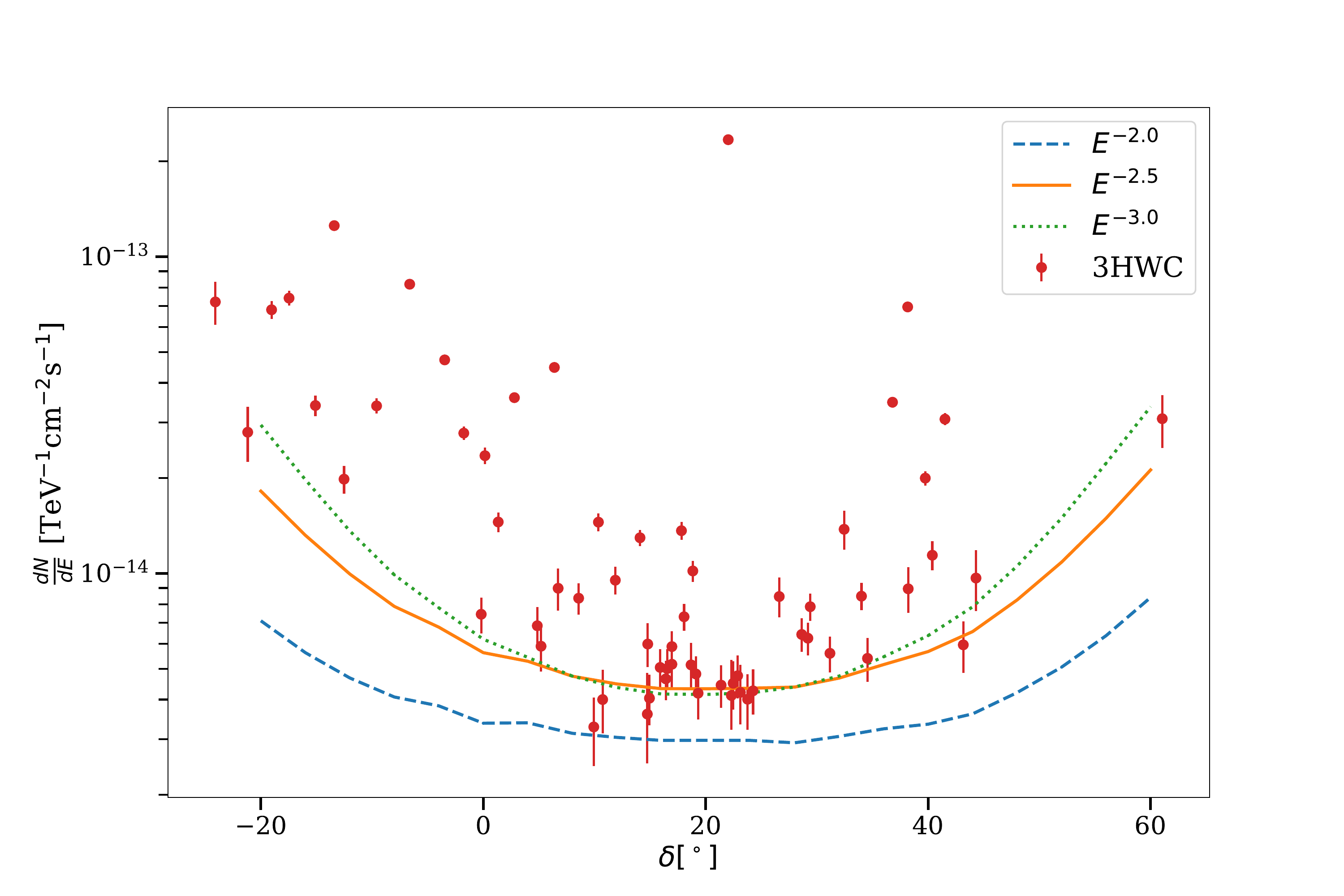}
\caption{3HWC sensitivity for the point source search as a function of declination. The flux sensitivity is shown at a pivot energy of 7 TeV for three spectral hypotheses: $E^{-2.0}$,$E^{-2.5}$ and $E^{-3.0}$. The sensitivity does not depend on the Right Ascension. Also shown is the best-fit flux normalization at 7 TeV for all sources in the 3HWC catalog.} \label{fig:sensi}
\end{figure}

We compare the flux measurements with the sensitivity for the underlying dataset. The flux sensitivity is defined as the flux normalization required to have a 50\% probability of detecting a source at the $5 \sigma$ level. Figure \ref{fig:sensi} shows the HAWC 1523-days sensitivity and the flux measurements from Table \ref{tab:fluxes} as a function of declination. HAWC is more sensitive to sources transiting directly overhead, corresponding to a declination of \ang{19.0}, than to sources transiting at larger zenith angles. HAWC is also more sensitive to hard-spectrum sources. For the optimal case (an $E^{-2}$ source transiting directly overhead), HAWC's sensitivity approaches $\sim 2\%$ of the flux of the Crab Nebula. The sensitivity is nearly constant with respect to the Right Ascension of a source (it varies by less than 3\% across the sky).

\begin{figure}[tbp]
\includegraphics[width=0.5\textwidth, trim=0cm 0cm 0cm 1.5cm, clip=true]{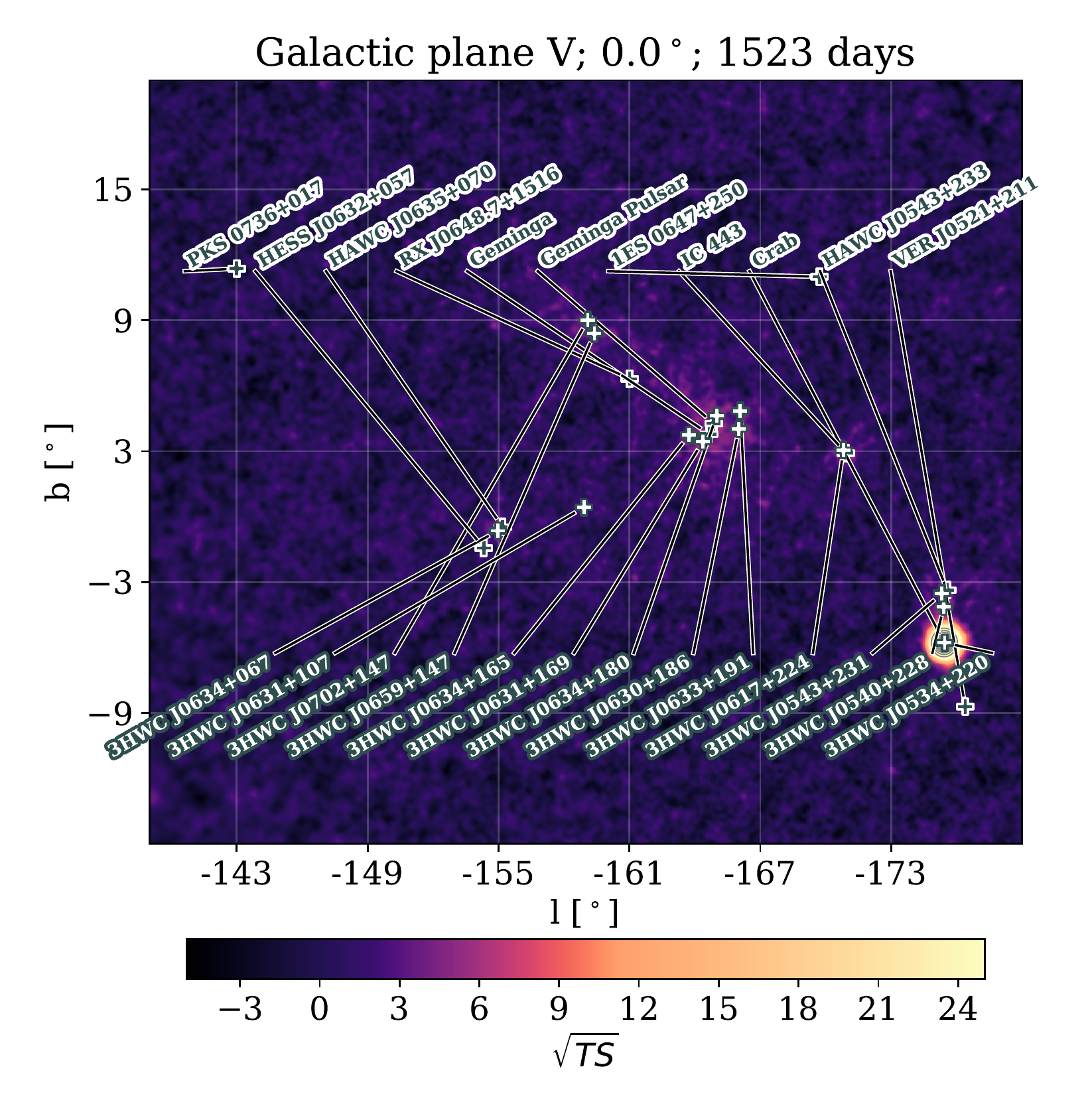}\includegraphics[width=0.5\textwidth, trim=0cm 0cm 0cm 1.5cm, clip=true]{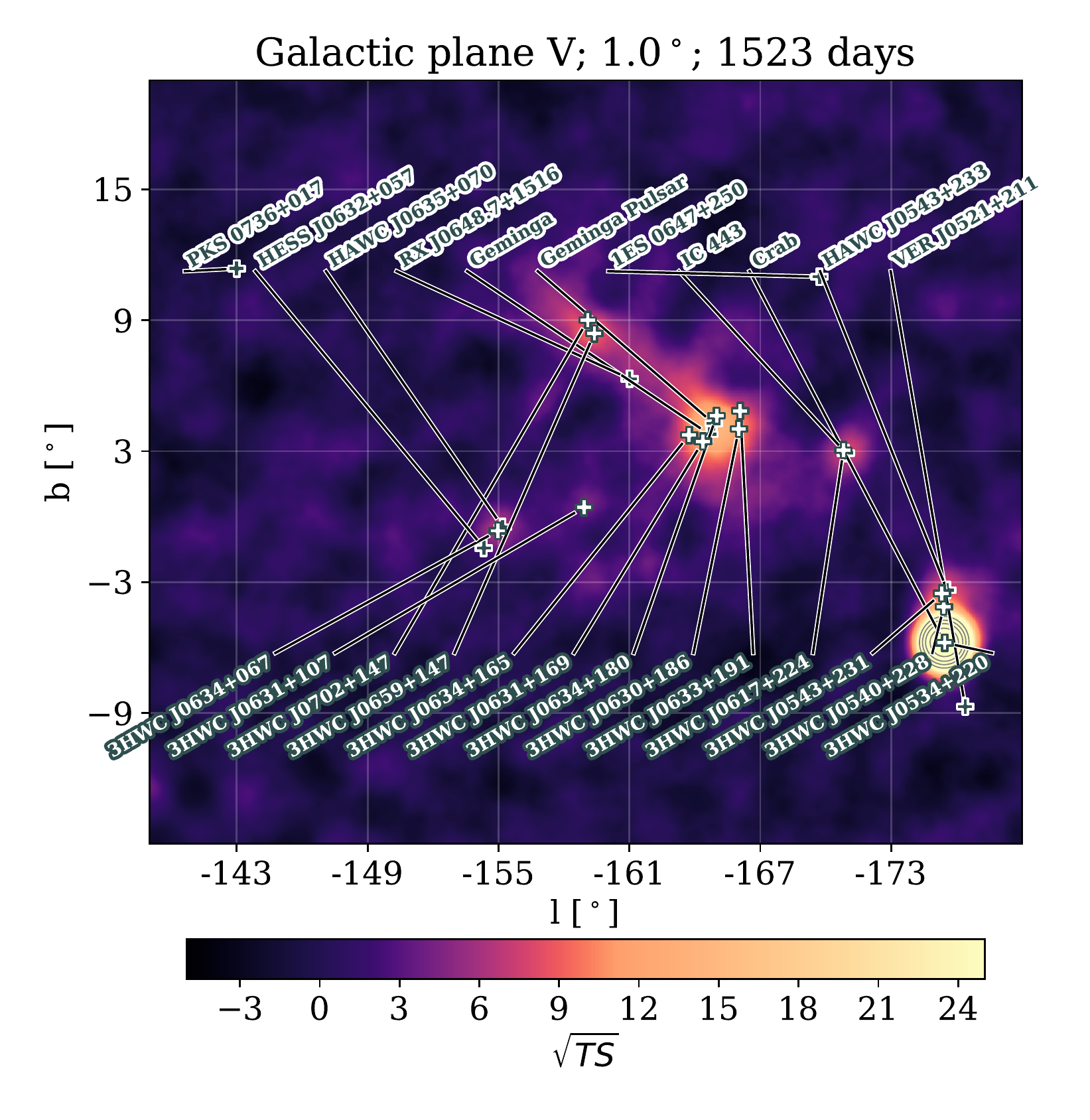}
\caption{Significance maps of the Galactic anti-center region for  $-176^\circ \leq l \leq -142^\circ$, showing the Crab Nebula and the Geminga halo among other sources. Left: Point-source hypothesis. Right: \ang{1} extended-source hypothesis. The grey lines show significance contours starting at $\sqrt{TS}=40$, increasing in steps of $\Delta \sqrt{TS}=20$.  Top labels indicate positions of known TeV sources (from TeVCat), bottom labels indicate positions of 3HWC sources.}
\label{fig:plane5}

\end{figure}

Most of the sources were found in the point source search. With about three times the livetime compared to the 2HWC catalog, many extended sources are now also significantly detected in the point source map. For example, Figure \ref{fig:plane5} shows five 3HWC sources (\textbf{3HWC J0630+186}, \textbf{3HWC J0631+169}, \textbf{3HWC J0633+191}, \textbf{3HWC J0634+165}, and \textbf{3HWC J0634+180}, all found in the point source search) clustering near the Geminga pulsar. We believe that these five sources are all part of the extended halo around Geminga, described in \cite{Abeysekara:2017old}. Similarly, both \textbf{3HWC J0659+147} and \textbf{3HWC J0702+147} are part of the extended source \textbf{2HWC J0700+143} announced in the aforementioned publication. It is not clear if these sources correspond to real features in the morphology of the two pulsar halos, or if they are just due to statistical fluctuations in the number of photons recorded by HAWC. 

\begin{figure}[t]
\includegraphics[width=0.9\textwidth, trim=0cm 0cm 0cm 1.5cm, clip=true]{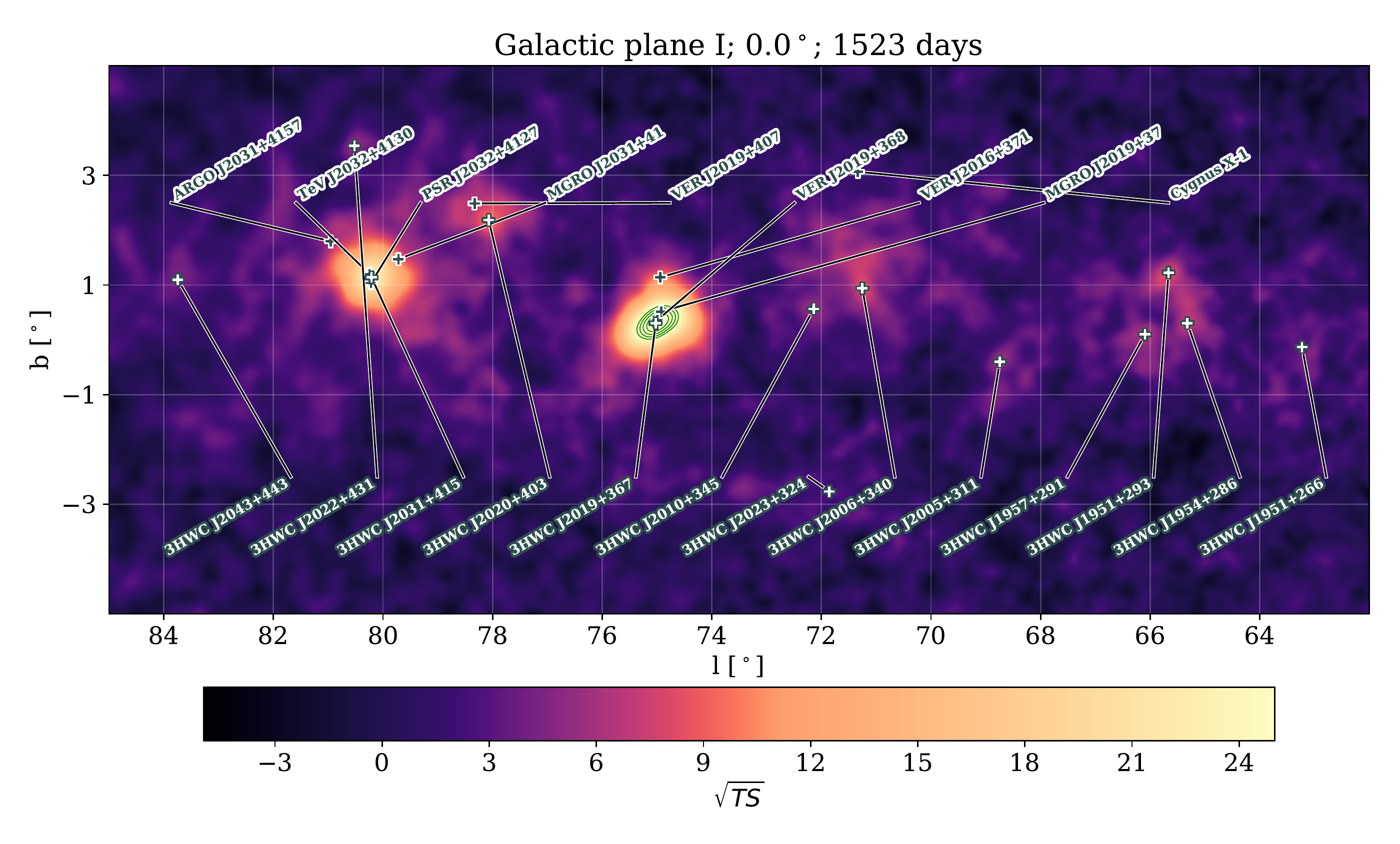}
\caption{Significance map of part of the Galactic plane for $62^\circ \leq l \leq 85^\circ$; point-source hypothesis. The green lines show significance contours starting at $\sqrt{TS}=26$, increasing in steps of $\Delta \sqrt{TS}=2$. Top labels indicate positions of known TeV sources (from TeVCat), bottom labels indicate positions of 3HWC sources.}
\label{fig:plane1}
\end{figure}

\begin{figure}[b]
\includegraphics[width=0.9\textwidth, trim=0cm 0cm 0cm 1.5cm, clip=true]{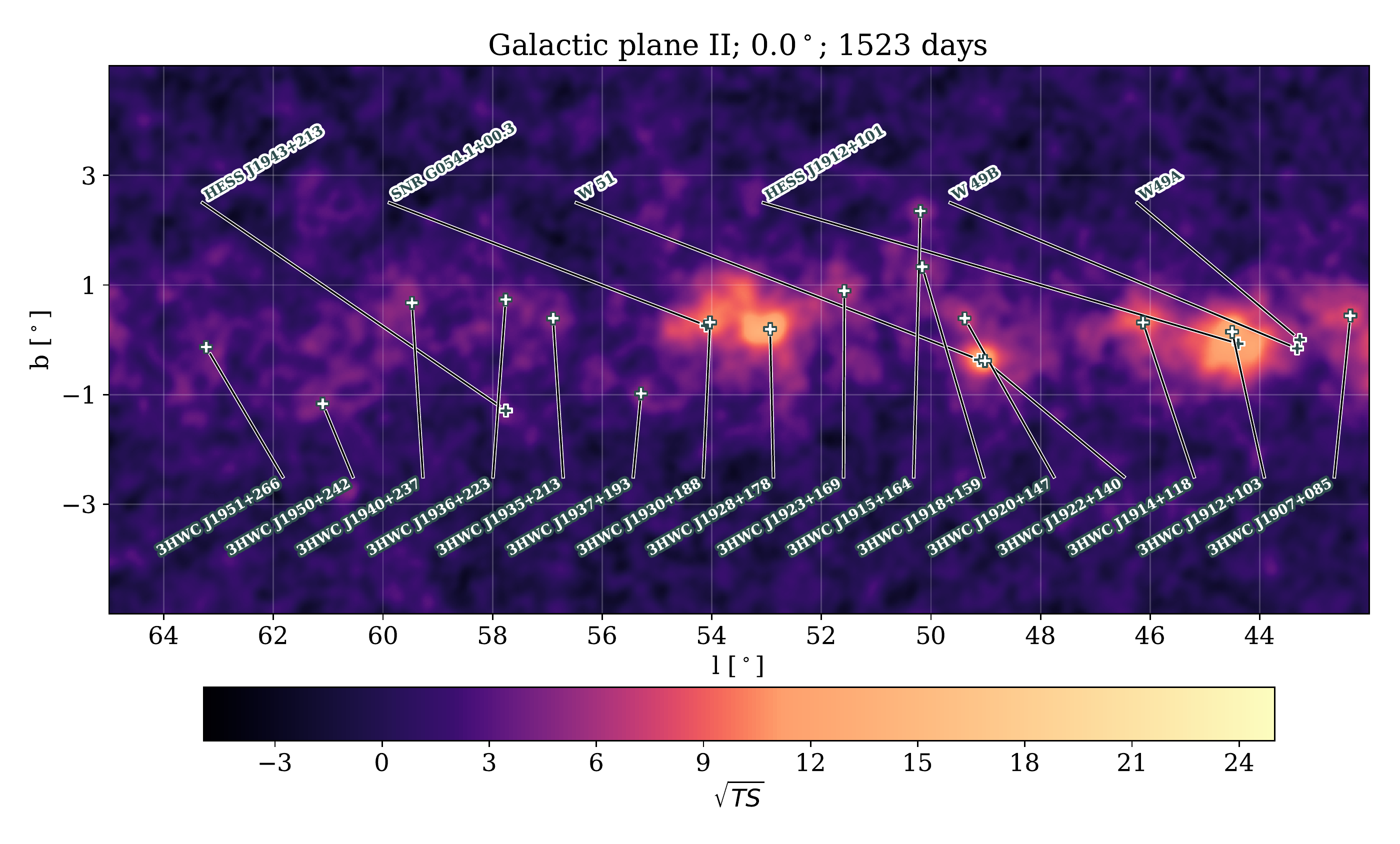}
\caption{Significance map of part of the Galactic plane for $42^\circ \leq l \leq 65^\circ$; point-source hypothesis. Top labels indicate positions of known TeV sources (from TeVCat), bottom labels indicate positions of 3HWC sources. }
\label{fig:plane2}
\end{figure}

\begin{figure}[t]
\includegraphics[width=0.9\textwidth, trim=0cm 0cm 0cm 1.5cm, clip=true]{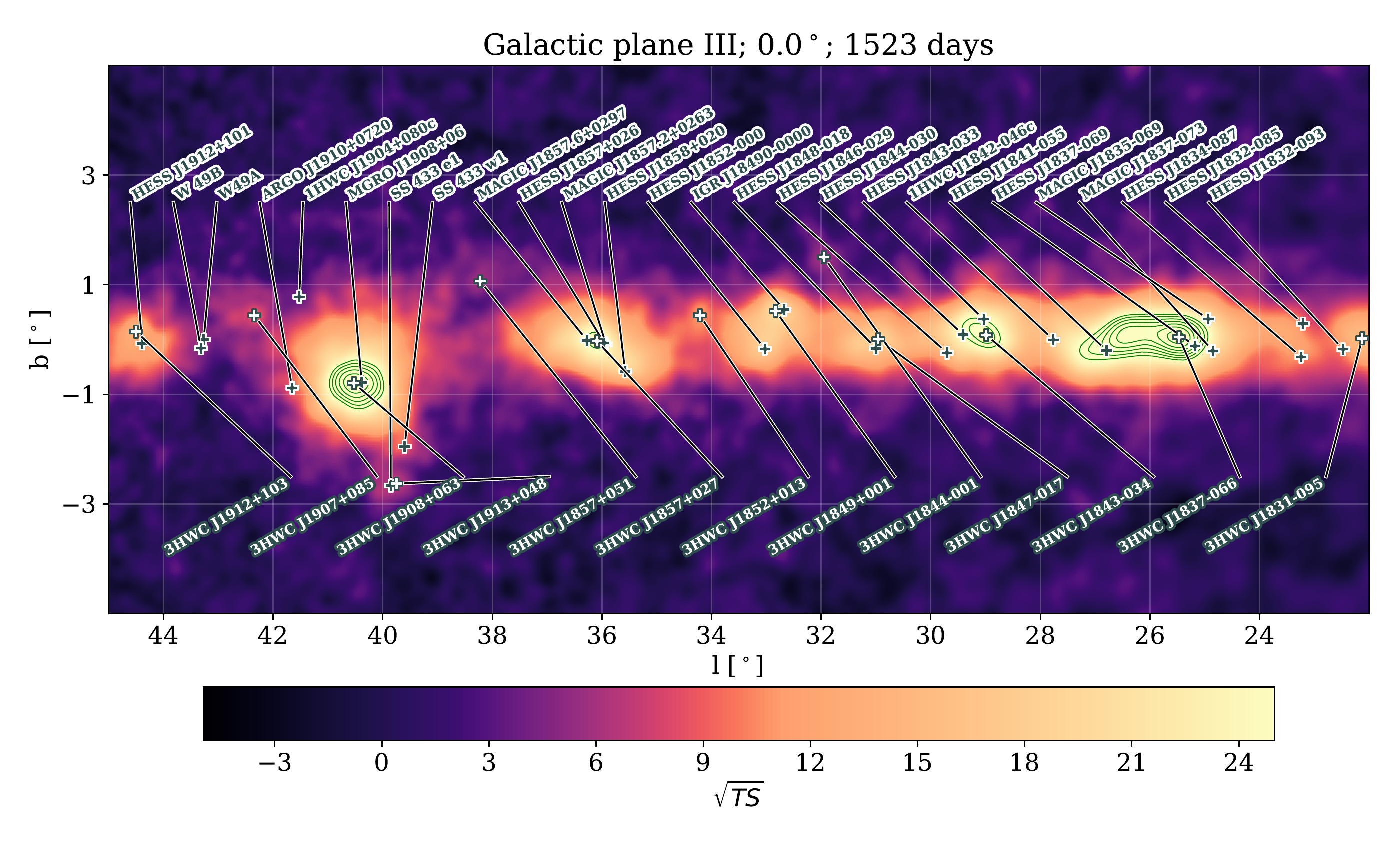}
\caption{Significance map of part of the Galactic plane for $22^\circ \leq l \leq 45^\circ$; point-source hypothesis. The green lines show significance contours starting at $\sqrt{TS}=26$, increasing in steps of $\Delta \sqrt{TS}=2$.  Top labels indicate positions of known TeV sources (from TeVCat), bottom labels indicate positions of 3HWC sources.}
\label{fig:plane3}
\end{figure}

\begin{figure}[b]
\includegraphics[width=0.9\textwidth, trim=0cm 0cm 0cm 1.5cm, clip=true]{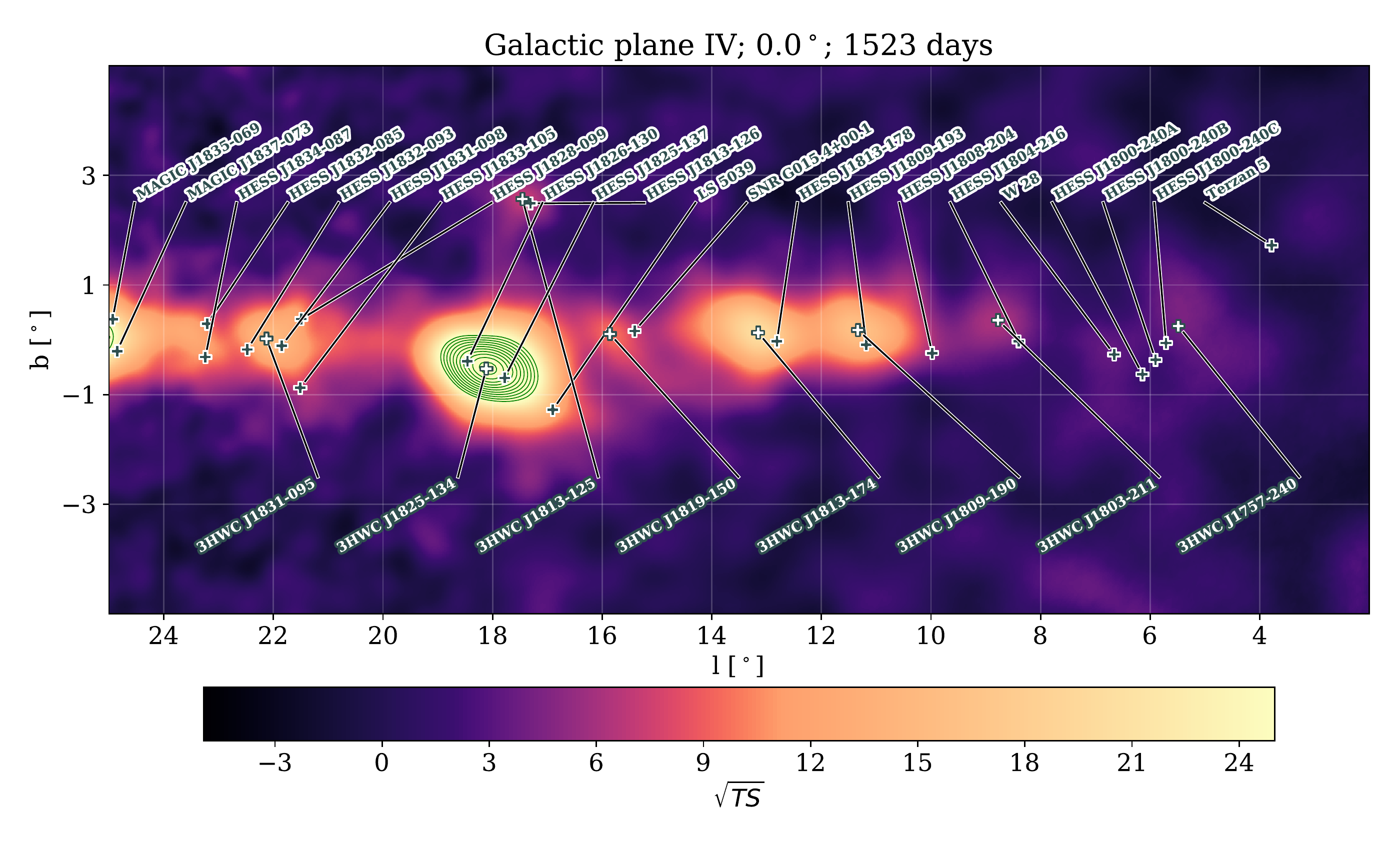}
\caption{Significance map of the inner Galactic plane for $2^\circ \leq l \leq 25^\circ$; point-source hypothesis. The green lines show significance contours starting at $\sqrt{TS}=26$, increasing in steps of $\Delta \sqrt{TS}=2$.  Top labels indicate positions of known TeV sources (from TeVCat), bottom labels indicate positions of 3HWC sources.}
\label{fig:plane4}
\end{figure}

\begin{figure}[tbp]
\plotone{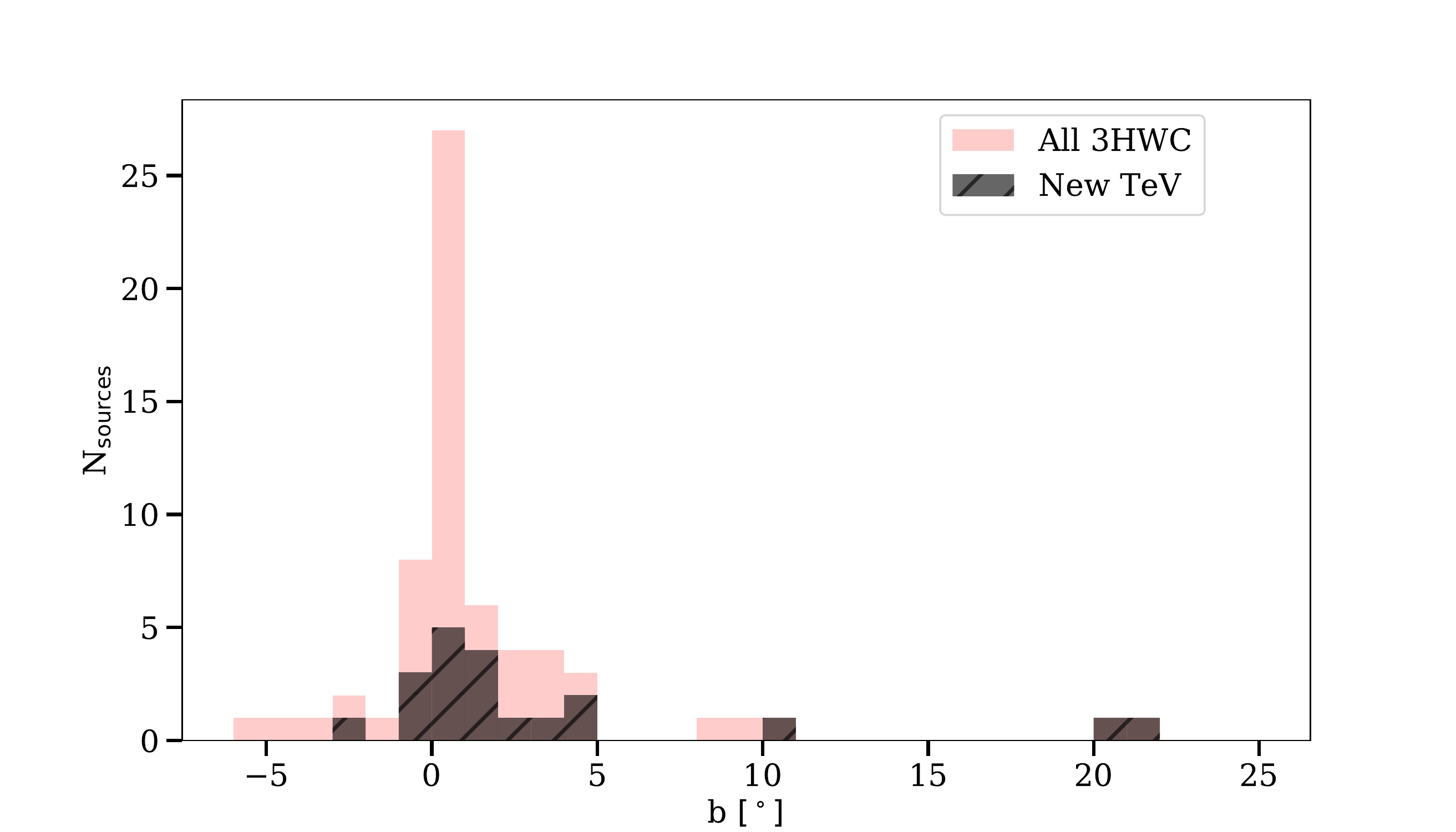}
\caption{Distribution of HAWC sources (excluding the known blazars) with respect to galactic latitude for \ang{-5} $< b <$ \ang{25}. The darker shaded histogram shows the new TeV sources in 3HWC that were not present in 2HWC. 
\label{fig:b_dist}}
\end{figure}

\begin{figure}[tbp]
\plotone{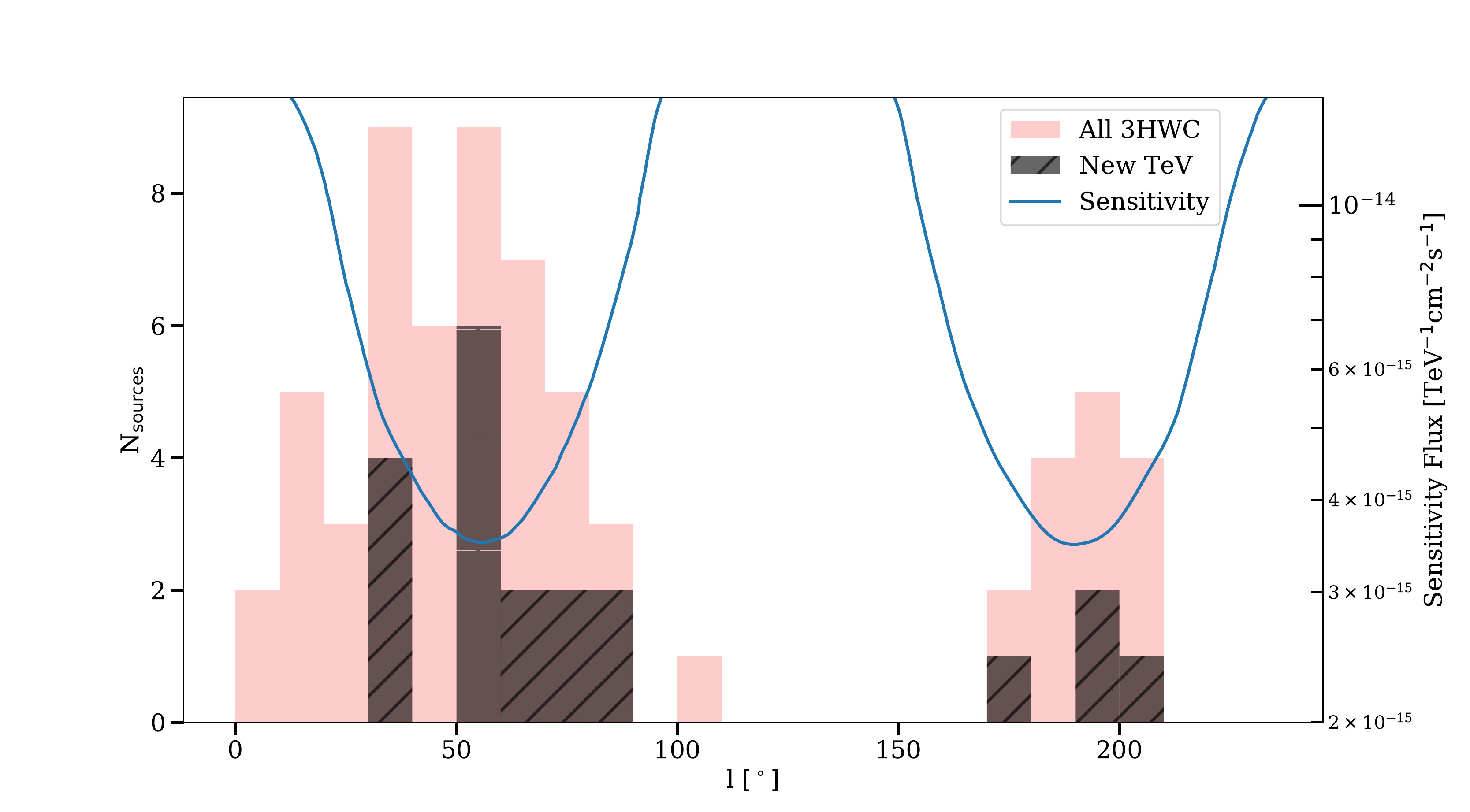}
\caption{Distribution of HAWC sources as a function of galactic longitude. The darker shaded histogram shows the new TeV sources in 3HWC that were not present in 2HWC. The blue solid line shows the sensitivity at $b =$  \ang{0}. Due to its location, HAWC is most sensitive towards the Galactic anticenter region, $l\approx$ \ang{-180} and, to the inner Galaxy at $l\approx +$\ang{50}. Most known TeV gamma-ray sources are located in the inner Galaxy.} \label{fig:l_dist}
\end{figure}

 As seen in the the all-sky significance map (Figure \ref{fig:allsky}), the majority of the sources in the 3HWC catalog are located along the Galactic plane. Figures \ref{fig:plane1}, \ref{fig:plane2}, \ref{fig:plane3}, and \ref{fig:plane4} show the significance maps of the Galactic plane from the Cygnus region (l=\ang{85}) to the inner Galaxy (l=\ang{2}). The Galactic center itself falls outside of the part of the sky visible to HAWC. Figure \ref{fig:plane5} shows a region near the Galactic anti-center containing the Crab Nebula, Geminga, and other sources. For this region, both the point-source significance map and the significance map from the \ang{1} extended source search are shown. For convenience, the locations of 3HWC sources and TeVCat sources have been marked in these images. Figures \ref{fig:b_dist} and \ref{fig:l_dist} show the distribution of 3HWC sources as a function of galactic latitude and longitude respectively.

\clearpage
 \startlongtable
  \begin{deluxetable*}{l c c c c c c c c c h h h}
    \tabletypesize{\small}
    \tablecaption{Source list and nearest TeVCat sources (within \ang{1} of each 3HWC source). Secondary sources (i.e., sources that are not separated from their
      neighbor(s) by a large TS gap) are marked with a dagger (\dag). The position uncertainty reported here is statistical only. The systematic uncertainty on the position is discussed in Section \ref{pointingbias}. TeVCat source names within \ang{0.5} of a 3HWC source are printed in bold. For sources without a TeVCat counterpart within \ang{1}, the angular distance to the nearest TeVCat source is printed for reference.
      \label{tab:sources}}
    \tablehead{
      \colhead{} & \colhead{} & \colhead{} & \colhead{} & \colhead{} & \colhead{} & \colhead{} & \colhead{} & \multicolumn{2}{c}{Nearest TeVCat source}\\
      \cline{9-10}
      \colhead{Name} & \colhead{Radius} & \colhead{TS} & \colhead{RA} & \colhead{Dec} & \colhead{\textit{l}} & \colhead{\textit{b}} & \colhead{1$\sigma$ stat. unc.} & \colhead{Dist.} & \colhead{Name} & \nocolhead{TeVCat flux} & \nocolhead{TeVCat index} & \nocolhead{TeVCat extent}\\
      \colhead{} & \colhead{[$^\circ$]} & \colhead{} & \colhead{[$^\circ$]} & \colhead{[$^\circ$]} & \colhead{[$^\circ$]} & \colhead{[$^\circ$]} & \colhead{[$^\circ$]} & \colhead{[$^\circ$]} & \colhead{} & \nocolhead{[CU]} & \nocolhead{} & \nocolhead{[$^\circ$]}
    }
    \startdata
       3HWC~J0534+220 &          0.0 &  35736.5 &  83.63 &  22.02 & 184.55 &  -5.78 &  0.06 &  0.01 &   \textbf{Crab} & 1.000 & -2.50 & 0.01 \\
3HWC~J0540+228$^{\dag}$ &          0.0 &     28.8 &  85.17 &  22.87 & 184.58 &  -4.13 &  0.11 &  0.77 &  HAWC J0543+233 &   N/A & -2.30 & 0.50 \\
 3HWC~J0543+231 &          0.0 &     34.2 &  85.78 &  23.11 & 184.67 &  -3.52 &  0.10 &  0.29 & \textbf{HAWC J0543+233} &   N/A & -2.30 & 0.50 \\
 3HWC~J0617+224 &          0.0 &     32.3 &  94.39 &  22.47 & 189.18 &   3.05 &  0.14 &  0.17 & \textbf{IC 443} & 0.030 & -3.00 & 0.16 \\
 3HWC~J0621+382 &          0.5 &     28.0 &  95.32 &  38.21 & 175.44 &  10.97 &  0.30 & 14.56 &             ... &   ... &   ... &    - \\
 3HWC~J0630+186 &          0.0 &     38.9 &  97.69 &  18.68 & 193.98 &   4.02 &  0.10 &  1.18 &             ... &   ... &   ... &    - \\
 3HWC~J0631+107 &          0.0 &     26.5 &  97.78 &  10.73 & 201.08 &   0.43 &  0.09 &  3.84 &             ... &   ... &   ... &    - \\
3HWC~J0631+169$^{\dag}$ &          0.0 &     39.1 &  97.95 &  16.96 & 195.63 &   3.45 &  0.17 &  0.44 & \textbf{Geminga} & 0.230 &   N/A & 2.60 \\
 3HWC~J0633+191 &          0.0 &     27.5 &  98.44 &  19.12 & 193.92 &   4.85 &  0.28 &  1.35 &             ... &   ... &   ... &    - \\
 3HWC~J0634+067 &          0.5 &     36.2 &  98.66 &   6.73 & 205.03 &  -0.65 &  0.22 &  0.28 & \textbf{HAWC J0635+070} &   N/A & -2.15 & 0.65 \\
3HWC~J0634+165$^{\dag}$ &          0.0 &     30.0 &  98.53 &  16.53 & 196.26 &   3.74 &  0.11 &  0.92 &         Geminga & 0.230 &   N/A & 2.60 \\
 3HWC~J0634+180 &          0.0 &     60.0 &  98.75 &  18.05 & 195.00 &   4.62 &  0.09 &  0.38 & \textbf{Geminga Pulsar} &   N/A &   N/A &   No \\
3HWC~J0659+147$^{\dag}$ &          0.0 &     28.8 & 104.85 &  14.75 & 200.60 &   8.40 &  0.12 &  0.50 & \textbf{2HWC J0700+143} &   N/A & -2.03 & 2.00 \\
 3HWC~J0702+147 &          0.0 &     28.9 & 105.56 &  14.75 & 200.91 &   9.01 &  0.19 &  0.60 &  2HWC J0700+143 &   N/A & -2.03 & 2.00 \\
 3HWC~J1104+381 &          0.0 &   3025.3 & 166.11 &  38.16 & 179.95 &  65.05 &  0.06 &  0.04 & \textbf{Markarian 421} & 0.300 & -2.20 &   No \\
 3HWC~J1654+397 &          0.0 &    227.8 & 253.52 &  39.74 &  63.58 &  38.82 &  0.09 &  0.05 & \textbf{Markarian 501} &   N/A & -2.72 &   No \\
 3HWC~J1739+099 &          0.0 &     28.2 & 264.99 &   9.93 &  33.89 &  20.34 &  0.06 &  2.87 &             ... &   ... &   ... &    - \\
 3HWC~J1743+149 &          0.0 &     25.9 & 265.82 &  14.94 &  39.13 &  21.68 &  0.10 &  4.61 &             ... &   ... &   ... &    - \\
 3HWC~J1757-240 &          1.0 &     28.6 & 269.30 & -24.09 &   5.49 &   0.25 &  0.48 &  0.74 & HESS J1800-240B &   N/A & -2.50 & 0.15 \\
3HWC~J1803-211$^{\dag}$ &          0.0 &     38.4 & 270.97 & -21.18 &   8.78 &   0.36 &  0.27 &  0.54 &  HESS J1804-216 & 0.250 &   N/A & 0.27 \\
 3HWC~J1809-190 &          0.0 &    264.8 & 272.46 & -19.04 &  11.33 &   0.18 &  0.08 &  0.31 & \textbf{HESS J1809-193} & 0.140 & -2.20 & 0.53 \\
 3HWC~J1813-125 &          0.0 &     51.9 & 273.34 & -12.52 &  17.46 &   2.57 &  0.14 &  0.17 & \textbf{HESS J1813-126} & 0.042 &   N/A & 0.21 \\
 3HWC~J1813-174 &          0.0 &    416.0 & 273.43 & -17.47 &  13.15 &   0.13 &  0.06 &  0.18 & \textbf{2HWC J1814-173} &   N/A & -2.61 &   No \\
3HWC~J1819-150$^{\dag}$ &          0.0 &     93.8 & 274.79 & -15.09 &  15.86 &   0.11 &  0.14 &  0.05 & \textbf{2HWC J1819-150*} &   N/A & -2.88 &   No \\
 3HWC~J1825-134 &          0.0 &   2212.5 & 276.46 & -13.40 &  18.12 &  -0.53 &  0.06 &  0.00 & \textbf{2HWC J1825-134} &   N/A & -2.58 &   No \\
 3HWC~J1831-095 &          0.0 &    237.7 & 277.87 &  -9.59 &  22.13 &   0.02 &  0.14 &  0.31 & \textbf{HESS J1831-098} & 0.040 & -2.10 & 0.15 \\
 3HWC~J1837-066 &          0.0 &   1542.7 & 279.40 &  -6.62 &  25.47 &   0.04 &  0.06 &  0.06 & \textbf{2HWC J1837-065} &   N/A & -2.90 &   No \\
 3HWC~J1843-034 &          0.0 &    876.6 & 280.99 &  -3.47 &  28.99 &   0.08 &  0.06 &  0.24 & \textbf{2HWC J1844-032} &   N/A & -2.64 &   No \\
3HWC~J1844-001$^{\dag}$ &          0.0 &     33.2 & 281.07 &  -0.19 &  31.95 &   1.50 &  0.15 &  1.20 &             ... &   ... &   ... &    - \\
 3HWC~J1847-017 &          0.0 &    338.1 & 281.95 &  -1.75 &  30.96 &   0.01 &  0.09 &  0.17 & \textbf{HESS J1848-018} & 0.020 & -2.80 & 0.32 \\
 3HWC~J1849+001 &          0.0 &    427.5 & 282.35 &   0.15 &  32.83 &   0.52 &  0.06 &  0.16 & \textbf{IGR J18490-0000} & 0.015 &   N/A &  Yes \\
3HWC~J1852+013$^{\dag}$ &          0.0 &    126.7 & 283.05 &   1.34 &  34.21 &   0.44 &  0.09 &  0.06 & \textbf{2HWC J1852+013*} &   N/A & -2.90 &   No \\
 3HWC~J1857+027 &          0.0 &    763.5 & 284.33 &   2.80 &  36.09 &  -0.03 &  0.06 &  0.14 & \textbf{HESS J1857+026} &   N/A & -2.39 & 0.11 \\
3HWC~J1857+051$^{\dag}$ &          0.0 &     35.3 & 284.33 &   5.19 &  38.22 &   1.06 &  0.11 &  1.23 &             ... &   ... &   ... &    - \\
 3HWC~J1907+085 &          0.0 &     75.5 & 286.79 &   8.57 &  42.35 &   0.44 &  0.09 &  0.07 & \textbf{2HWC J1907+084*} &   N/A & -3.25 &   No \\
 3HWC~J1908+063 &          0.0 &   1320.9 & 287.05 &   6.39 &  40.53 &  -0.80 &  0.06 &  0.14 & \textbf{MGRO J1908+06} & 0.170 & -2.10 & 0.34 \\
 3HWC~J1912+103 &          0.0 &    198.2 & 288.06 &  10.35 &  44.50 &   0.15 &  0.09 &  0.24 & \textbf{HESS J1912+101} & 0.100 & -2.70 & 0.26 \\
3HWC~J1913+048$^{\dag}$ &          0.0 &     44.7 & 288.32 &   4.86 &  39.75 &  -2.63 &  0.13 &  0.11 & \textbf{SS 433 e1} &   N/A & -2.00 & 0.35 \\
 3HWC~J1914+118 &          0.0 &    102.9 & 288.68 &  11.87 &  46.13 &   0.32 &  0.09 &  0.15 & \textbf{2HWC J1914+117*} &   N/A & -2.83 &   No \\
 3HWC~J1915+164 &          0.0 &     27.5 & 288.76 &  16.41 &  50.19 &   2.35 &  0.10 &  2.92 &             ... &   ... &   ... &    - \\
 3HWC~J1918+159 &          0.0 &     31.6 & 289.69 &  15.91 &  50.16 &   1.33 &  0.11 &  1.99 &             ... &   ... &   ... &    - \\
3HWC~J1920+147$^{\dag}$ &          0.0 &     55.4 & 290.17 &  14.79 &  49.39 &   0.39 &  0.10 &  0.80 &            W 51 & 0.030 &   N/A & 0.12 \\
 3HWC~J1922+140 &          0.0 &    176.6 & 290.70 &  14.09 &  49.01 &  -0.38 &  0.06 &  0.10 &   \textbf{W 51} & 0.030 &   N/A & 0.12 \\
3HWC~J1923+169$^{\dag}$ &          0.0 &     46.1 & 290.79 &  16.96 &  51.58 &   0.89 &  0.10 &  1.54 &             ... &   ... &   ... &    - \\
 3HWC~J1928+178 &          0.0 &    216.7 & 292.10 &  17.82 &  52.93 &   0.20 &  0.08 &  0.06 & \textbf{2HWC J1928+177} &   N/A & -2.56 &   No \\
 3HWC~J1930+188 &          0.0 &    115.6 & 292.54 &  18.84 &  54.03 &   0.32 &  0.09 &  0.09 & \textbf{SNR G054.1+00.3} & 0.025 & -2.39 &   No \\
3HWC~J1935+213$^{\dag}$ &          0.0 &     26.2 & 293.95 &  21.38 &  56.90 &   0.39 &  0.12 &  1.89 &             ... &   ... &   ... &    - \\
 3HWC~J1936+223 &          0.0 &     28.2 & 294.08 &  22.31 &  57.76 &   0.73 &  0.11 &  1.62 &             ... &   ... &   ... &    - \\
 3HWC~J1937+193 &          0.0 &     25.2 & 294.39 &  19.31 &  55.29 &  -0.98 &  0.13 &  1.72 &             ... &   ... &   ... &    - \\
 3HWC~J1940+237 &          0.0 &     27.2 & 295.05 &  23.77 &  59.47 &   0.67 &  0.36 &  0.28 & \textbf{2HWC J1938+238} &   N/A & -2.96 &   No \\
 3HWC~J1950+242 &          0.0 &     25.1 & 297.69 &  24.26 &  61.10 &  -1.16 &  0.11 &  0.32 & \textbf{2HWC J1949+244} &   N/A & -2.38 & 1.00 \\
3HWC~J1951+266$^{\dag}$ &          0.5 &     35.6 & 297.90 &  26.61 &  63.23 &  -0.13 &  0.61 &  2.14 &             ... &   ... &   ... &    - \\
 3HWC~J1951+293 &          0.0 &     68.7 & 297.99 &  29.40 &  65.66 &   1.23 &  0.14 &  0.25 & \textbf{2HWC J1953+294} &   N/A & -2.78 &   No \\
3HWC~J1954+286$^{\dag}$ &          0.0 &     48.3 & 298.70 &  28.63 &  65.32 &   0.30 &  0.14 &  0.12 & \textbf{2HWC J1955+285} &   N/A & -2.40 &   No \\
 3HWC~J1957+291 &          0.0 &     41.0 & 299.36 &  29.18 &  66.09 &   0.10 &  0.10 &  0.75 &  2HWC J1955+285 &   N/A & -2.40 &   No \\
 3HWC~J2005+311 &          0.0 &     33.1 & 301.46 &  31.17 &  68.74 &  -0.40 &  0.12 &  3.01 &             ... &   ... &   ... &    - \\
 3HWC~J2006+340 &          0.0 &     67.4 & 301.73 &  34.00 &  71.25 &   0.94 &  0.14 &  0.23 & \textbf{2HWC J2006+341} &   N/A & -2.64 &   No \\
3HWC~J2010+345$^{\dag}$ &          0.0 &     27.6 & 302.69 &  34.55 &  72.14 &   0.56 &  0.17 &  1.01 &             ... &   ... &   ... &    - \\
 3HWC~J2019+367 &          0.0 &   1227.5 & 304.94 &  36.80 &  75.02 &   0.30 &  0.06 &  0.07 & \textbf{VER J2019+368} &   N/A & -1.75 & 0.34 \\
 3HWC~J2020+403 &          0.0 &     93.9 & 305.16 &  40.37 &  78.07 &   2.19 &  0.09 &  0.40 & \textbf{VER J2019+407} & 0.037 & -2.37 & 0.23 \\
 3HWC~J2022+431 &          0.0 &     29.0 & 305.52 &  43.16 &  80.52 &   3.54 &  0.10 &  1.80 &             ... &   ... &   ... &    - \\
 3HWC~J2023+324 &          1.0 &     30.7 & 305.81 &  32.44 &  71.85 &  -2.77 &  0.73 &  3.96 &             ... &   ... &   ... &    - \\
 3HWC~J2031+415 &          0.0 &    556.9 & 307.93 &  41.51 &  80.21 &   1.14 &  0.06 &  0.07 & \textbf{TeV J2032+4130} & 0.030 & -2.00 & 0.19 \\
3HWC~J2043+443$^{\dag}$ &          0.5 &     28.6 & 310.89 &  44.30 &  83.74 &   1.10 &  0.24 &  2.88 &             ... &   ... &   ... &    - \\
 3HWC~J2227+610 &          0.0 &     52.5 & 336.96 &  61.05 & 106.42 &   2.87 &  0.19 &  0.16 & \textbf{Boomerang} & 0.440 &   N/A &  Yes \\

    \enddata
  \end{deluxetable*}

\clearpage 

  \startlongtable
   \begin{deluxetable*}{l c c c c}
     \tabletypesize{\small}
    \tablecaption{Source radius, best-fit spectrum, and
      energy range.
      The flux $F_7$ is the differential flux at 7\,TeV.
      The two sets of reported uncertainties correspond to statistical and systematic, respectively.
        The spectral fit for \textbf{3HWC J0659+147} did not converge.   The energy range quoted here is the true energy interval from which we expect to get 75\% of a given source's significance.
      \label{tab:fluxes}}
    \tablehead{
      \colhead{Name} & \colhead{Radius} & \colhead{Index} & \multicolumn{1}{c}{$F_{7}$} & \colhead{Energy Range}\\
      \colhead{} & \colhead{[$^\circ$]} & \colhead{} & \multicolumn{1}{c}{[$10^{-15}$ TeV$^{-1}$cm$^{-2}$s$^{-1}$]} & \colhead{[TeV]}
    }
    \startdata
      \decimals
       3HWC~J0534+220 & 0.0 &         $-2.579$ $\pm 0.005$ $ ^{+0.067}_{-0.018}$ &                234.2 $\pm 1.4$ $ ^{+55.2}_{-34.0}$ & 1.6 -- 37.4\\
 3HWC~J0540+228 & 0.0 &             $-2.84$ $\pm 0.14$ $ ^{+0.14}_{-0.03}$ &             4.8 $^{+0.7}_{-0.8}$ $ ^{+1.1}_{-0.8}$ & 1.1 -- 28.5\\
 3HWC~J0543+231 & 0.0 &     $-2.13$ $^{+0.15}_{-0.16}$ $ ^{+0.16}_{-0.01}$ &                    4.2 $\pm 0.9$ $ ^{+1.9}_{-0.6}$ & 10.5 -- 205.9\\
 3HWC~J0617+224 & 0.0 &             $-3.05$ $\pm 0.11$ $ ^{+0.06}_{-0.02}$ &                    4.5 $\pm 0.8$ $ ^{+1.2}_{-0.8}$ & 0.5 -- 12.0\\
 3HWC~J0621+382 & 0.5 &     $-2.41$ $^{+0.12}_{-0.13}$ $ ^{+0.08}_{-0.01}$ &             8.9 $^{+1.4}_{-1.5}$ $ ^{+2.6}_{-1.2}$ & 5.7 -- 138.3\\
 3HWC~J0630+186 & 0.0 &     $-2.21$ $^{+0.13}_{-0.14}$ $ ^{+0.14}_{-0.02}$ &             5.1 $^{+0.8}_{-0.9}$ $ ^{+2.1}_{-0.8}$ & 8.4 -- 183.8\\
 3HWC~J0631+107 & 0.0 &     $-2.23$ $^{+0.17}_{-0.19}$ $ ^{+0.10}_{-0.01}$ &             4.0 $^{+0.9}_{-1.0}$ $ ^{+1.4}_{-0.6}$ & 8.6 -- 186.9\\
 3HWC~J0631+169 & 0.0 &     $-2.51$ $^{+0.13}_{-0.14}$ $ ^{+0.13}_{-0.03}$ &                    5.9 $\pm 0.7$ $ ^{+2.0}_{-0.9}$ & 3.3 -- 95.0\\
 3HWC~J0633+191 & 0.0 &     $-2.64$ $^{+0.14}_{-0.15}$ $ ^{+0.10}_{-0.02}$ &                    4.8 $\pm 0.7$ $ ^{+1.3}_{-0.7}$ & 2.1 -- 61.9\\
 3HWC~J0634+067 & 0.5 &             $-2.27$ $\pm 0.10$ $ ^{+0.09}_{-0.01}$ &             9.0 $^{+1.3}_{-1.4}$ $ ^{+2.7}_{-1.2}$ & 7.6 -- 171.1\\
 3HWC~J0634+165 & 0.0 &     $-2.52$ $^{+0.19}_{-0.22}$ $ ^{+0.13}_{-0.03}$ &             5.0 $^{+0.7}_{-0.8}$ $ ^{+1.6}_{-0.8}$ & 3.2 -- 93.0\\
 3HWC~J0634+180 & 0.0 &     $-2.47$ $^{+0.10}_{-0.11}$ $ ^{+0.11}_{-0.02}$ &                    7.3 $\pm 0.7$ $ ^{+2.2}_{-1.1}$ & 3.7 -- 102.0\\
 3HWC~J0659+147 & 0.0 &                                                ... &                                                ... & ...\\
 3HWC~J0702+147 & 0.0 &     $-1.99$ $^{+0.19}_{-0.24}$ $ ^{+0.18}_{-0.02}$ &             3.6 $^{+1.1}_{-1.3}$ $ ^{+1.9}_{-0.5}$ & 14.1 -- 261.8\\
 3HWC~J1104+381 & 0.0 &             $-3.04$ $\pm 0.01$ $ ^{+0.06}_{-0.02}$ &                 69.4 $\pm 1.3$ $ ^{+16.6}_{-11.4}$ & 0.7 -- 15.0\\
 3HWC~J1654+397 & 0.0 &             $-2.91$ $\pm 0.04$ $ ^{+0.06}_{-0.02}$ &                   20.0 $\pm 1.0$ $ ^{+5.3}_{-3.2}$ & 1.3 -- 29.4\\
 3HWC~J1739+099 & 0.0 &     $-1.98$ $^{+0.13}_{-0.14}$ $ ^{+0.12}_{-0.02}$ &                    3.3 $\pm 0.8$ $ ^{+1.4}_{-0.6}$ & 14.8 -- 274.0\\
 3HWC~J1743+149 & 0.0 &     $-2.37$ $^{+0.15}_{-0.16}$ $ ^{+0.08}_{-0.02}$ &             4.0 $^{+0.7}_{-0.8}$ $ ^{+1.3}_{-0.6}$ & 5.6 -- 139.3\\
 3HWC~J1757-240 & 1.0 &             $-2.80$ $\pm 0.11$ $ ^{+0.07}_{-0.03}$ &         72.0 $^{+10.9}_{-11.3}$ $ ^{+17.6}_{-9.2}$ & 5.6 -- 158.4\\
 3HWC~J1803-211 & 0.0 &     $-2.59$ $^{+0.13}_{-0.14}$ $ ^{+0.09}_{-0.04}$ &           27.9 $^{+5.4}_{-5.7}$ $ ^{+10.3}_{-4.4}$ & 9.7 -- 206.8\\
 3HWC~J1809-190 & 0.0 &             $-2.59$ $\pm 0.05$ $ ^{+0.09}_{-0.03}$ &           68.0 $^{+4.4}_{-4.5}$ $ ^{+24.5}_{-9.7}$ & 7.7 -- 177.3\\
 3HWC~J1813-125 & 0.0 &             $-2.81$ $\pm 0.10$ $ ^{+0.09}_{-0.03}$ &                   19.9 $\pm 2.0$ $ ^{+5.9}_{-3.1}$ & 2.6 -- 69.2\\
 3HWC~J1813-174 & 0.0 &             $-2.54$ $\pm 0.04$ $ ^{+0.09}_{-0.02}$ &                 74.0 $\pm 4.0$ $ ^{+26.7}_{-10.7}$ & 7.7 -- 174.8\\
 3HWC~J1819-150 & 0.0 &     $-2.90$ $^{+0.08}_{-0.07}$ $ ^{+0.10}_{-0.03}$ &                  33.9 $\pm 2.5$ $ ^{+10.3}_{-5.4}$ & 2.2 -- 62.5\\
 3HWC~J1825-134 & 0.0 &             $-2.35$ $\pm 0.02$ $ ^{+0.11}_{-0.02}$ &                125.4 $\pm 3.4$ $ ^{+53.1}_{-18.3}$ & 9.2 -- 183.4\\
 3HWC~J1831-095 & 0.0 &             $-2.61$ $\pm 0.06$ $ ^{+0.12}_{-0.03}$ &           33.8 $^{+1.8}_{-1.9}$ $ ^{+12.2}_{-5.3}$ & 4.2 -- 106.7\\
 3HWC~J1837-066 & 0.0 &             $-2.73$ $\pm 0.02$ $ ^{+0.11}_{-0.03}$ &                 81.9 $\pm 1.7$ $ ^{+25.2}_{-13.2}$ & 2.2 -- 57.3\\
 3HWC~J1843-034 & 0.0 &             $-2.36$ $\pm 0.03$ $ ^{+0.14}_{-0.02}$ &                  47.2 $\pm 1.6$ $ ^{+19.6}_{-7.2}$ & 6.2 -- 142.6\\
 3HWC~J1844-001 & 0.0 &             $-2.76$ $\pm 0.12$ $ ^{+0.12}_{-0.03}$ &                    7.4 $\pm 1.0$ $ ^{+2.4}_{-1.3}$ & 1.9 -- 51.2\\
 3HWC~J1847-017 & 0.0 &             $-2.87$ $\pm 0.04$ $ ^{+0.10}_{-0.04}$ &                   27.7 $\pm 1.3$ $ ^{+8.4}_{-5.1}$ & 1.4 -- 33.1\\
 3HWC~J1849+001 & 0.0 &             $-2.17$ $\pm 0.04$ $ ^{+0.16}_{-0.01}$ &                  23.5 $\pm 1.4$ $ ^{+11.6}_{-3.9}$ & 9.9 -- 195.3\\
 3HWC~J1852+013 & 0.0 &             $-2.79$ $\pm 0.08$ $ ^{+0.13}_{-0.04}$ &            14.5 $^{+1.0}_{-1.1}$ $ ^{+4.4}_{-2.6}$ & 1.7 -- 40.2\\
 3HWC~J1857+027 & 0.0 &             $-2.83$ $\pm 0.03$ $ ^{+0.10}_{-0.03}$ &                  35.9 $\pm 1.2$ $ ^{+10.2}_{-6.5}$ & 1.3 -- 31.8\\
 3HWC~J1857+051 & 0.0 &             $-3.03$ $\pm 0.12$ $ ^{+0.09}_{-0.04}$ &                    5.9 $\pm 1.0$ $ ^{+1.8}_{-1.2}$ & 0.7 -- 15.5\\
 3HWC~J1907+085 & 0.0 &             $-2.95$ $\pm 0.09$ $ ^{+0.09}_{-0.04}$ &             8.4 $^{+0.9}_{-1.0}$ $ ^{+2.5}_{-1.6}$ & 0.8 -- 19.7\\
 3HWC~J1908+063 & 0.0 &             $-2.12$ $\pm 0.03$ $ ^{+0.18}_{-0.02}$ &                  44.7 $\pm 1.4$ $ ^{+20.7}_{-7.1}$ & 8.9 -- 182.7\\
 3HWC~J1912+103 & 0.0 &             $-2.85$ $\pm 0.05$ $ ^{+0.09}_{-0.03}$ &                   14.5 $\pm 0.9$ $ ^{+4.4}_{-2.6}$ & 1.1 -- 27.1\\
 3HWC~J1913+048 & 0.0 &             $-2.37$ $\pm 0.13$ $ ^{+0.14}_{-0.03}$ &             6.8 $^{+0.9}_{-1.0}$ $ ^{+2.8}_{-1.1}$ & 5.9 -- 144.1\\
 3HWC~J1914+118 & 0.0 &                        $-2.9$ $\pm 0.1$ $ \pm 0.1$ &             9.5 $^{+0.9}_{-1.0}$ $ ^{+2.9}_{-1.9}$ & 1.0 -- 23.5\\
 3HWC~J1915+164 & 0.0 &             $-2.60$ $\pm 0.13$ $ ^{+0.07}_{-0.01}$ &                    4.6 $\pm 0.7$ $ ^{+1.4}_{-0.7}$ & 2.5 -- 69.5\\
 3HWC~J1918+159 & 0.0 &     $-2.49$ $^{+0.15}_{-0.17}$ $ ^{+0.13}_{-0.02}$ &                    5.1 $\pm 0.7$ $ ^{+1.8}_{-0.8}$ & 3.6 -- 101.9\\
 3HWC~J1920+147 & 0.0 &             $-2.98$ $\pm 0.13$ $ ^{+0.11}_{-0.05}$ &             6.0 $^{+0.9}_{-1.0}$ $ ^{+2.0}_{-1.3}$ & 0.7 -- 16.8\\
 3HWC~J1922+140 & 0.0 &             $-2.62$ $\pm 0.06$ $ ^{+0.09}_{-0.02}$ &                   13.0 $\pm 0.8$ $ ^{+3.7}_{-2.0}$ & 2.1 -- 60.0\\
 3HWC~J1923+169 & 0.0 &             $-3.07$ $\pm 0.10$ $ ^{+0.07}_{-0.03}$ &                    5.2 $\pm 0.8$ $ ^{+1.4}_{-0.9}$ & 0.5 -- 11.1\\
 3HWC~J1928+178 & 0.0 &             $-2.30$ $\pm 0.07$ $ ^{+0.17}_{-0.02}$ &                   13.6 $\pm 0.9$ $ ^{+5.3}_{-2.1}$ & 5.9 -- 140.5\\
 3HWC~J1930+188 & 0.0 &             $-2.76$ $\pm 0.07$ $ ^{+0.09}_{-0.03}$ &                   10.2 $\pm 0.8$ $ ^{+2.9}_{-1.7}$ & 1.4 -- 36.6\\
 3HWC~J1935+213 & 0.0 &             $-2.73$ $\pm 0.16$ $ ^{+0.08}_{-0.03}$ &                    4.4 $\pm 0.7$ $ ^{+1.3}_{-0.7}$ & 1.6 -- 43.0\\
 3HWC~J1936+223 & 0.0 &                 $-2.9$ $\pm 0.3$ $ ^{+0.2}_{-0.1}$ &             4.1 $^{+0.9}_{-1.2}$ $ ^{+1.5}_{-1.2}$ & 0.9 -- 20.9\\
 3HWC~J1937+193 & 0.0 &             $-2.90$ $\pm 0.16$ $ ^{+0.09}_{-0.04}$ &             4.2 $^{+0.7}_{-0.8}$ $ ^{+1.2}_{-0.8}$ & 0.9 -- 22.2\\
 3HWC~J1940+237 & 0.0 &     $-3.14$ $^{+0.12}_{-0.11}$ $ ^{+0.07}_{-0.03}$ &                    4.0 $\pm 0.8$ $ ^{+1.1}_{-0.8}$ & 0.4 -- 7.9\\
 3HWC~J1950+242 & 0.0 &     $-2.49$ $^{+0.18}_{-0.19}$ $ ^{+0.15}_{-0.02}$ &                    4.3 $\pm 0.7$ $ ^{+1.5}_{-0.7}$ & 3.7 -- 102.2\\
 3HWC~J1951+266 & 0.5 &     $-2.36$ $^{+0.12}_{-0.13}$ $ ^{+0.11}_{-0.02}$ &             8.5 $^{+1.2}_{-1.3}$ $ ^{+2.6}_{-1.3}$ & 5.5 -- 133.7\\
 3HWC~J1951+293 & 0.0 &     $-2.47$ $^{+0.09}_{-0.10}$ $ ^{+0.10}_{-0.02}$ &                    7.9 $\pm 0.8$ $ ^{+2.7}_{-1.2}$ & 4.0 -- 108.6\\
 3HWC~J1954+286 & 0.0 &             $-2.42$ $\pm 0.12$ $ ^{+0.15}_{-0.02}$ &                    6.4 $\pm 0.8$ $ ^{+2.4}_{-1.0}$ & 4.8 -- 119.7\\
 3HWC~J1957+291 & 0.0 &     $-2.54$ $^{+0.13}_{-0.14}$ $ ^{+0.11}_{-0.02}$ &                    6.2 $\pm 0.7$ $ ^{+2.1}_{-1.0}$ & 3.3 -- 93.9\\
 3HWC~J2005+311 & 0.0 &             $-2.58$ $\pm 0.13$ $ ^{+0.11}_{-0.02}$ &                    5.6 $\pm 0.7$ $ ^{+1.8}_{-0.9}$ & 3.0 -- 83.6\\
 3HWC~J2006+340 & 0.0 &     $-2.56$ $^{+0.12}_{-0.13}$ $ ^{+0.15}_{-0.03}$ &             8.5 $^{+0.8}_{-0.9}$ $ ^{+3.1}_{-1.5}$ & 3.3 -- 83.5\\
 3HWC~J2010+345 & 0.0 &             $-2.91$ $\pm 0.13$ $ ^{+0.10}_{-0.03}$ &                    5.4 $\pm 0.9$ $ ^{+1.5}_{-0.9}$ & 1.1 -- 24.3\\
 3HWC~J2019+367 & 0.0 &             $-2.04$ $\pm 0.02$ $ ^{+0.16}_{-0.02}$ &                  34.7 $\pm 1.3$ $ ^{+17.8}_{-5.7}$ & 11.7 -- 211.7\\
 3HWC~J2020+403 & 0.0 &             $-3.11$ $\pm 0.07$ $ ^{+0.08}_{-0.04}$ &                   11.4 $\pm 1.2$ $ ^{+3.4}_{-2.2}$ & 0.7 -- 14.8\\
 3HWC~J2022+431 & 0.0 &             $-2.34$ $\pm 0.12$ $ ^{+0.10}_{-0.02}$ &                    6.0 $\pm 1.1$ $ ^{+2.2}_{-0.9}$ & 8.3 -- 181.9\\
 3HWC~J2023+324 & 1.0 &     $-2.70$ $^{+0.13}_{-0.12}$ $ ^{+0.06}_{-0.03}$ &            13.8 $^{+1.9}_{-2.0}$ $ ^{+3.2}_{-2.1}$ & 2.0 -- 52.9\\
 3HWC~J2031+415 & 0.0 &             $-2.36$ $\pm 0.04$ $ ^{+0.14}_{-0.02}$ &           30.7 $^{+1.3}_{-1.4}$ $ ^{+13.1}_{-4.9}$ & 6.3 -- 147.8\\
 3HWC~J2043+443 & 0.5 &     $-2.33$ $^{+0.14}_{-0.15}$ $ ^{+0.11}_{-0.02}$ &             9.7 $^{+2.1}_{-2.2}$ $ ^{+3.9}_{-1.5}$ & 8.5 -- 185.7\\
 3HWC~J2227+610 & 0.0 &             $-2.43$ $\pm 0.10$ $ ^{+0.10}_{-0.02}$ &           30.8 $^{+5.9}_{-5.8}$ $ ^{+13.1}_{-4.1}$ & 14.3 -- 292.7\\

    \enddata
  \end{deluxetable*}%

\clearpage

\clearpage

\clearpage

\subsection{Comparison with the 2HWC catalog}
Thirty-three of the 40 the sources detected in the 2HWC catalog have a 3HWC counterpart within \ang{1}. Most of HAWC's sources are supernova remnants, pulsar wind nebula, or pulsar halos, and are expected to have constant emission, with the exception of the two Markarians, which are known to be variable at TeV energies on time scales of hours to weeks \citep[e.g.][]{Abeysekara_2017_Mrk}. For these sources, the TS increased by a factor of 2.3, on average. This is slightly less than the expected improvement due to the increase in livetime (factor of 3). The apparent deficit is explained by the new instrument response functions and the change in spectral index that was used in the source searches, verified by re-running the catalog search with the new settings on the 2HWC dataset.

There are seven 2HWC sources without a 3HWC counterpart within \ang{1}. Two of these sources (\textbf{2HWC J1902+048} and \textbf{2HWC J2024+417}) still show significant emission ($TS>25$) in the new data set, but do not pass the TS dip test. \textbf{2HWC J1902+048} is now considered part of the \textbf{3HWC~J1908+063} complex, and \textbf{2HWC J2024+417} is part of the \textbf{3HWC~J2031+415} complex.

The remaining five sources do not pass the $TS>25$ criterion with the new data set. Out of these, \textbf{2HWC J1921+131} lies in the Galactic plane near the W51 supernova remnant. In the 3HWC data set (point-source search), \textbf{2HWC J1921+131} has a TS of 21.7, which is below the threshold for significant detection. It is possible that the 2HWC detection was due to an upward fluctuation of a combination of diffuse emission and emission from the nearby W51 complex. The other four sources (\textbf{2HWC J0819+157}, \textbf{2HWC J1040+308}, \textbf{2HWC J1309-054}, and \textbf{2HWC J1829+070}) were located off the Galactic plane and had no known counterparts. Due to the low detection significance of these sources, searches for variability of these sources on time-scales of several months have not yielded conclusive results. (We note that these searches were mainly focused on a change in detection significance scaled over time. A more detailed analysis of the light-curves of these sources will follow in future HAWC publications.) It is unclear whether the previous detections were false positives due to random background fluctuations or upward fluctuations of real, but weak, sources, or indicative of flaring activity/temporal variability. We estimate the expected number of false positives, i.e. random background fluctuations that pass the detection threshold, to be 0.75 for the 3HWC catalog (see Section \ref{sec:discuss}), compared to 0.4 for 2HWC.

Four of the seven 2HWC sources without a 3HWC equivalent (\textbf{2HWC~J0819+157}, \textbf{2HWC~J1040+308}, \textbf{2HWC~J1829+070}, and \textbf{2HWC~J1902+048}) are also not detected anymore when applying the new search methods/settings to the 2HWC dataset.

\subsection{New 3HWC Sources Potentially Associated with Known TeV Sources}
Eight of the 3HWC sources have no counterpart (within \ang{1}) in 2HWC, but are potentially associated with known TeV emitters. These sources are listed below. We define positional coincidence within \ang{1} as the criterion for two sources to be considered as candidates for association. 

\textbf{3HWC~J0540+228} ($TS=28.8$) and \textbf{3HWC~J0543+231} ($TS=34.2$) are part of the extended source that had been previously announced as \textbf{HAWC~J0543+233} \citep{2017ATel10941....1R} -- a potential TeV halo around the pulsar \textbf{PSR~B0540+23}.

\textbf{3HWC~J0617+224} ($TS=32.3$) is potentially associated with the shell-type supernova remnant (SNR) \textbf{IC~443} (\textbf{SNR~G189.1+03.0}), which has been detected at TeV energies by MAGIC \citep{2007ApJ...664L..87A} and VERITAS \citep{2009ApJ...698L.133A}. HAWC previously announced the detection of gamma-ray emission from \textbf{IC~443} without naming the source \citep{Fleischhack:2019njo}.

\textbf{3HWC~J0634+067} ($TS=36.2$) had been previously announced as \textbf{HAWC~J0635+070} \citep{2018ATel12013....1B} -- a potential TeV halo around the pulsar \textbf{PSR~J0633+0632}.

\textbf{3HWC~J2227+610} ($TS=52.5$) is within \ang{1} of the known TeV gamma-ray sources \textbf{VER~J2227+608} \citep{2009ApJ...703L...6A} and \textbf{MGRO~J2228+61} \citep{Abdo_2009_MGRO,Abdo_2009_erratum}. The most likely source of the emission is the shell-type supernova remnant \textbf{SNR~G106.3+2.7}. \textbf{3HWC~J2227+610} was recently announced as \textbf{HAWC~J2227+610} in a dedicated publication \citep{boomerangPaper}.

\textbf{3HWC~J1913+048} ($TS=44.7$) is within \ang{1} of the eastern lobe of the micro-quasar \textbf{SS~433}. Detection of TeV gamma-ray emission from this source (as well as the western lobe of \textbf{SS~433}) had previously been announced by HAWC \citep{Abeysekara:2018qtj}.

\textbf{3HWC J1757-240} ($TS=28.6$) was found in the \ang{1} extended source search. It overlaps with the W28 region, which contains at least four known TeV sources (\textbf{HESS J1801-233}, and \textbf{HESS J1800-240A/B/C}) \citep{2008A&A...481..401A}. The emission seen by H.E.S.S. has been attributed to an SNR interacting with nearby molecular clouds. Due to HAWC's limited sensitivity and relatively poor angular resolution at these declinations, we are currently unable to resolve the individual H.E.S.S. sources.

\textbf{3HWC J1803-211} ($TS=38.4$) is located near the known, but unidentified, TeV source \textbf{HESS J1804-216} \citep{2005Sci...307.1938A,2006ApJ...636..777A}. 

\subsection{New TeV Sources}
\label{sec:new_sources}
For each source in 3HWC, we scan several catalogs of known or potential gamma-ray sources for potential associations within \ang{1} including the TeVCat \citep{2008ICRC....3.1341W},  the fourth \textit{Fermi}-LAT source catalog (4FGL) \citep{Fermi-LAT:2019yla}, the ATNF Pulsar Catalog \footnote{\url{https://www.atnf.csiro.au/research/pulsar/psrcat/}} (v 1.62) \citep{Manchester:2004bp}, and the Galactic supernova remnant catalog SNRCat \footnote{\url{http://www.physics.umanitoba.ca/snr/SNRcat}} \citep{2012AdSpR..49.1313F}.. 

We report 20 new sources that do not have a potential counterpart in the TeVCat (see Table \ref{tab:gev}). Fourteen of these new sources are within \ang{1} of a previously observed GeV source. Table \ref{tab:gev} lists the GeV associations and their source classifications obtained from the fourth \textit{Fermi}-LAT source catalog, 4FGL. Two new sources, \textbf{3HWC J0630+186} ($TS=38.9$) and \textbf{3HWC J1918+159} ($TS=31.6$), have no GeV counterpart in 4FGL. These two sources are, however, potentially associated with pulsars in the ATNF catalog. \textbf{3HWC J0630+186} is within \ang{0.95} of the pulsar \textbf{PSR J0630+19}. \textbf{3HWC J1918+159} is potentially associated with \textbf{PSR J1918+1541} with a separation distance of \ang{0.26}. The age and spin-down luminosity of these objects are not available. 

\begin{deluxetable}{c c c c c c c c}
\tabletypesize{\small}
\tablecaption{New HAWC sources with no TeV counterpart. For each source we list the following information in the various columns: Galactic longitude; Galactic latitude; the nearest GeV source in 4FGL \citep{Fermi-LAT:2019yla} and its separation from the 3HWC source; the source class as listed in 4FGL where available (bcu: active galaxy of uncertain type, PSR: pulsar, identified by pulsations, unk: unknown); the nearest pulsar and corresponding separation from the ATNF pulsar catalog \citep{Manchester:2004bp}; and the nearest SNR, separation distance, and type from the SNRCat \citep{2012AdSpR..49.1313F}.}
\label{tab:gev}

\tablehead{
\colhead{HAWC} & \colhead{\textit{l} [$^\circ$]} & \colhead{\textit{b} [$^\circ$]} & \colhead{4FGL ($^\circ$)} & \colhead{Class} & \colhead{ATNF ($^\circ$)} & \colhead{SNRCat ($^\circ$)} & \colhead{SNR Type}
}

\startdata
3HWC J0621+382 & 175.44 & 10.97 & 4FGL J0620.3+3804  (0.22)  & bcu   & J0622+3749 (0.42) &  ... & ... \\ 
3HWC J0630+186 & 193.98 & 4.02 &   ... & ... & J0630+19 (0.94) &  ... & \\
3HWC J0631+107 & 201.08 & 0.43 & 4FGL J0631.5+1036  (0.15)  & PSR   & J0631+1036 (0.14) &  ... & ... \\
3HWC J0633+191 & 193.92 & 4.85 &    ... & ... &  ... &  ... & ... \\
3HWC J1739+099 & 33.89 & 20.34 & 4FGL J1740.5+1005  (0.22)  & PSR   & J1740+1000 (0.13) & G034.0+20.3 (0.13) & filled-centre\\
3HWC J1743+149 & 39.13 & 21.68 &    ... & ... &  ... &  ... & \\
3HWC J1844$-$001 & 31.95 & 1.50 & 4FGL J1848.2$-$0016  (0.99)  &       ... & J1843$-$0000 (0.27) &  ... & ... \\
3HWC J1857+051 & 38.22 & 1.06 & 4FGL J1855.2+0456  (0.56)  &       ... & J1857+0526 (0.24) &  ... & ... \\
3HWC J1915+164 & 50.19 & 2.35 & 4FGL J1912.0+1612  (0.74)  & bcu   & B1913+16 (0.32) &  ... & ... \\
3HWC J1918+159 & 50.16 & 1.33 &   ... & ... & J1918+1541 (0.26) &  ... & ... \\
3HWC J1923+169 & 51.58 & 0.89 & 4FGL J1925.1+1707  (0.50)  & unk   & B1921+17 (0.14) &  ... & ... \\
3HWC J1935+213 & 56.90 & 0.39 & 4FGL J1935.2+2029  (0.89)  & PSR   & J1936+21 (0.24) & G057.2+00.8 (0.59) & shell\\
3HWC J1936+223 & 57.76 & 0.73 & 4FGL J1932.2+2221  (0.94)  & PSR   & J1938+2213 (0.44) & G057.2+00.8 (0.47) & shell\\
3HWC J1937+193 & 55.29 & $-$0.98 & 4FGL J1936.6+1921  (0.21)  &      ... & J1936+20 (0.77) &  ... & ... \\
3HWC J1951+266 & 63.23 & $-$0.13 & 4FGL J1951.6+2621  (0.25)  &      ... & J1952+2630 (0.24) &  ... & ... \\ 
3HWC J2005+311 & 68.74 & $-$0.40 & 4FGL J2006.2+3102  (0.15)  & PSR   & J2006+3102 (0.15) & G068.6$-$01.2 (0.81) & unknown\\
3HWC J2010+345 & 72.14 & 0.56 &   ... & ... &  ... &  ... & ... \\
3HWC J2022+431 & 80.52 & 3.54 &   ... & ... &  ... &  ... & ... \\
3HWC J2023+324 & 71.85 & $-$2.77 & 4FGL J2024.0+3202  (0.43)  & unk   &  ... &  ... & ... \\
3HWC J2043+443 & 83.74 & 1.10 & 4FGL J2047.5+4356  (0.79)  &       ... &  ... &  ... & ... \\
\enddata
\end{deluxetable}

\subsubsection{Unassociated New TeV Sources}
We observe four sources that do not have an apparent counterpart in any of the catalogs that we scanned for potential associations: \textbf{3HWC J0633+191} ($TS=37.5$), \textbf{3HWC J2010+345} ($TS=27.6$), \textbf{3HWC J2022+431} ($TS=29.0$), and \textbf{3HWC J1743+149} ($TS=25.9$). We note that three of these new sources are not well isolated from known extended sources in the catalog. 
\textbf{3HWC J0633+191} is in a dense region of known pulsars including Geminga.
\textbf{3HWC J2010+345} is near \textbf{3HWC J2004+343}, which itself is a 1$^\circ$ extended source. 
\textbf{3HWC J2022+431} is in the Cygnus-X region with a number of star-forming clusters nearby, most notably the \textit{Fermi}-LAT cocoon \citep{Hona:2019ysf}. Without a detailed morphological study of their respective regions, we cannot definitively exclude the above new sources as appendages of existing sources. Such a study, however, is beyond the scope of this paper.

\textbf{3HWC J1743+149} (TS = 25.9) is the only new unassociated source that is not in spatial proximity of a region of known TeV sources. It is also notably distant from the Galactic plane with a Galactic latitude of  $b = $ \ang{21.7}. The nearest potential GeV gamma-ray counterpart is 4FGL~J1741.4+1354, associated to the pulsar PSR~J1741+1351, at a distance of \ang{1.3} from 3HWC~J1743+149 (outside of our nominal search radius for counterparts).

\subsubsection{Pulsars and TeV Halo Candidates in the 3HWC}

\begin{figure}[tb]
\plotone{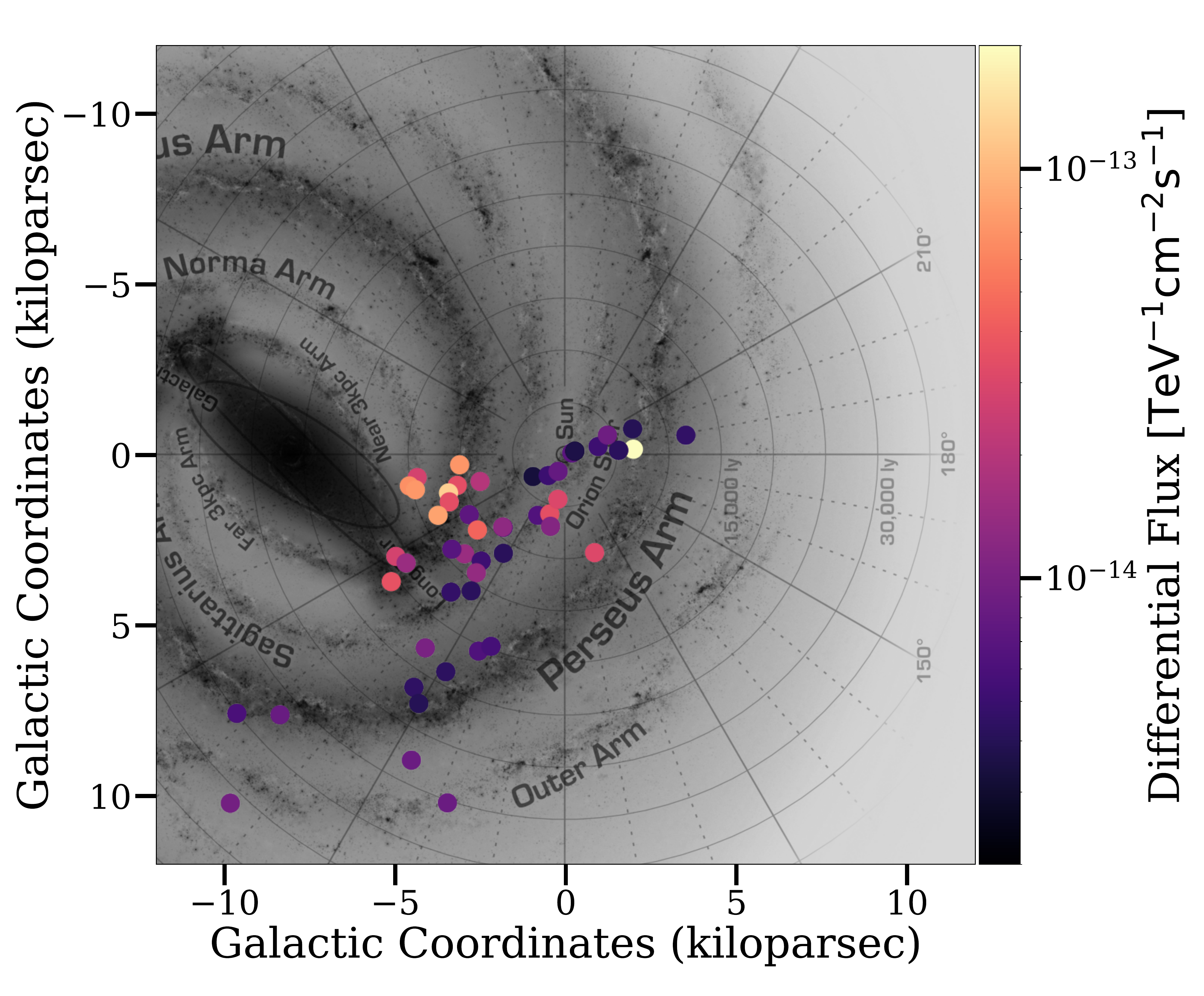}
\caption{Face-on view of the galaxy showing positions of HAWC sources associated with (i.e., spatially coincident within 1$^\circ$ of) pulsars for which distances are estimated. Spatial coincidence does not necessarily imply that the observed gamma-ray emission is (fully) powered by the pulsar in question. The color scale corresponds to the measured flux normalization from Table \ref{tab:fluxes}. The annotated Milky Way background is taken from \cite{mw_pic}.
\label{fig:mw_dist}}
\end{figure}

A significant fraction of 3HWC sources are candidates for association with pulsars in the ATNF catalog (c.f. Table \ref{tab:gev}). Figure \ref{fig:mw_dist} shows all the 3HWC sources potentially associated with pulsars in the galaxy for which the distance information is available. 


After the discovery of extended gamma-ray emission around the Geminga and Monogem pulsars by HAWC \citep{Abeysekara:2017old}, and the discovery of several extended TeV PWNe by H.E.S.S. \citep{Abdalla:2017vci}, it has been suggested that extended pulsar ``Halos'' are a common feature or such objects and that several unassociated/not firmly associated HAWC sources may be dominated by such Halo emission \citep[see e.g.][]{PhysRevD.96.103016,Linden:2017blp,Giacinti:2019nbu,PhysRevD.100.043016,Fleischhack:2019rjb,DiMauro:2019yvh,Manconi:2020ipm,PhysRevD.101.103035}. The observed gamma-ray emission is thought to be due to inverse Compton up-scattering of cosmic microwave background photons by relativistic electrons and positrons diffusing freely in the vicinity around the pulsar. TeV halos are thought to form around older pulsars (at least several tens of thousands of years old) that have either left their SNR shell or whose SNR shell has already dissipated. They are thus distinct from (classical) PWNe, where the electron-positron plasma is confined by the ambient medium.

We produce an updated list of pulsars that are likely candidates to have a TeV Halo detectable with HAWC, following similar criteria as \cite{PhysRevD.96.103016}. We select pulsars from the ATNF with ages between 100\,kyr and 400\,kyr, declinations between \ang{-25} and +\ang{64}, and an estimated spindown flux of at least 1\% of that of the Geminga pulsar. We find sixteen such pulsars, out of which eight are coincident with at least one 3HWC source (within \ang{1}). Table \ref{tab:psrhalo} lists the 3HWC sources that are coincident with these TeV Halo candidate pulsars. Some pulsars have more than one 3HWC source nearby. This is not unexpected as our source search sometimes finds multiple point sources associated with the same extended emission region. One of these pulsars, \textbf{PSR J1740+1000}, has not previously been detected at TeV energies.


\begin{deluxetable}{c c c c c c c c c}
\tabletypesize{\small}
\tablecaption{HAWC Sources with the corresponding TeV halo candidate pulsars within $1^\circ$. The age of the pulsar in kyr and the spin-down luminosity, $\dot{E}$, in erg s$^{-1}$ are also given. The Separation column indicates the angular distance between the HAWC source and the ATNF pulsar \citep{Manchester:2004bp}. The TeVCat column lists the previously detected TeV counterpart of each source.}
\label{tab:psrhalo}
 \tablehead{
\colhead{HAWC }& \colhead{\textit{l} [$^\circ$]} & \colhead{\textit{b} [$^\circ$]} & \colhead{Pulsar} & \colhead{Age [kyr]} &\colhead{$\dot{E}$ [erg s$^{-1}$]} & \colhead{Distance [kpc]} & \colhead{Separation [$^\circ$]} & \colhead{TeVCat}
}

\startdata
3HWC J0540+228 & 184.58 & -4.13 & B0540+23 & 253.0 & 4.09e+34 & 1.56 & 0.83 & HAWC J0543+233\\
3HWC J0543+231 & 184.67 & -3.52 & B0540+23 & 253.0 & 4.09e+34 & 1.56 & 0.36 & HAWC J0543+233\\
3HWC J0631+169 & 195.63 & 3.45 & J0633+1746 & 342.0 & 3.25e+34 & 0.19 & 0.95 & Geminga\\
3HWC J0634+180 & 195.00 & 4.62 & J0633+1746 & 342.0 & 3.25e+34 & 0.19 & 0.38 & Geminga Pulsar\\
3HWC J0659+147 & 200.60 & 8.40 & B0656+14 & 111.0 & 3.8e+34 & 0.29 & 0.51 & 2HWC J0700+143\\
3HWC J0702+147 & 200.91 & 9.01 & B0656+14 & 111.0 & 3.8e+34 & 0.29 & 0.77 & 2HWC J0700+143\\
3HWC J1739+099 & 33.89 & 20.34 & J1740+1000 & 114.0 & 2.32e+35 & 1.23 & 0.13 & ... \\
3HWC J1831-095 & 22.13 & 0.02 & J1831$-$0952 & 128.0 & 1.08e+36 & 3.68 & 0.27 & HESS J1831-098\\
3HWC J1912+103 & 44.50 & 0.15 & J1913+1011 & 169.0 & 2.87e+36 & 4.61 & 0.31 & HESS J1912+101\\
3HWC J1923+169 & 51.58 & 0.89 & J1925+1720 & 115.0 & 9.54e+35 & 5.06 & 0.67 & ... \\
3HWC J1928+178 & 52.93 & 0.20 & J1925+1720 & 115.0 & 9.54e+35 & 5.06 & 0.85 & 2HWC J1928+177\\
3HWC J2031+415 & 80.21 & 1.14 & J2032+4127 & 201.0 & 1.52e+35 & 1.33 & 0.11 & TeV J2032+4130\\
\enddata

\end{deluxetable}

\section{Limitations and Systematic Uncertainties}\label{sec:discuss}

\subsection{Background Fluctuations and Spurious Detections}
It is possible for mere fluctuations in the background and/or the Galactic diffuse emission to pass the selection criteria and produce a spurious source. In order to estimate the frequency of false positive sources, we create twenty simulated significance maps using the background counts from the original source search. For each map, we obtain the simulated number of signal events in each pixel by Poisson-fluctuating the number of background events in the corresponding pixel. We then run each of these randomized background maps through the full analysis pipeline, including point and extended searches. In the 20 total randomized background maps, we find 15 local maxima with a $TS > 25$. Therefore, the estimated number of false positive sources is $15/20 = 0.75$. The fluctuations are distributed evenly across the sky and typically occur just above the threshold value of $TS = 25$.

\subsection{Limitations of the Source Search}
As in the 2HWC catalog, we conduct blind source searches for four different fixed morphological assumptions (point sources, and \ang{0.5}, \ang{1.0}, \ang{2.0} extended sources). We then combine these results, with preference given to sources found in the point source search and the smaller radius searches to avoid double counting of sources.

This approach can lead to sources being misidentified or missed. First, some extended sources may be significant enough to be detected in the point source analysis. Poisson fluctuations of the signal could potentially lead to several hotspots being detected around the center of such an extended source. As HAWC collects more data, this issue is increasing in prevalence, as evidenced by the five point sources detected inside the Geminga halo. Second, it is also possible that multiple smaller sources located near each other are detected as one source in the extended source search if the individual sources are not strong enough to cross the detection threshold. This might be happening near the Galactic center where \textbf{3HWC J1757-240}, found in the \ang{1} extended-source search, overlaps with several known TeV sources. Third, weaker sources may be missed if they are located near a stronger source, as they may not produce a well-defined peak in the significance map.  In-depth studies (such as \cite{Abeysekara:2017old, Abeysekara:2018qtj}) are needed to properly resolve source-dense regions. Such studies include multi-source fits, fitting the extensions/shapes, locations, and spectra of several sources the same time. Additionally, measurements by other gamma-ray observatories as well as measurements at other wavelengths might help disentangle the morphology of complex regions. Further studies of selected regions of the Galactic plane are in preparation.

\subsection{Systematic Uncertainty on the Source Locations}\label{pointingbias}

Earlier publications \citep{2HWC} quoted an \ang{0.1} systematic pointing uncertainty, which was estimated using simulations and verified through the observation of the Crab Nebula, Mrk421, and Mrk501. New studies of HAWC's pointing calibration as a function of source position suggest that the uncertainty could be larger than previously thought for sources that transit near the edge of HAWC's field of view. HAWC's absolute pointing uncertainty increases to \ang{0.15} for sources at \ang{-10} or +\ang{50} declination and could be as high as \ang{0.3} at declinations of \ang{-20} or +\ang{60}. (There are no well-isolated point sources detected by HAWC that could be used to unambiguously verify the instrument's pointing at these declinations.)

\begin{figure}[tbp]
\plotone{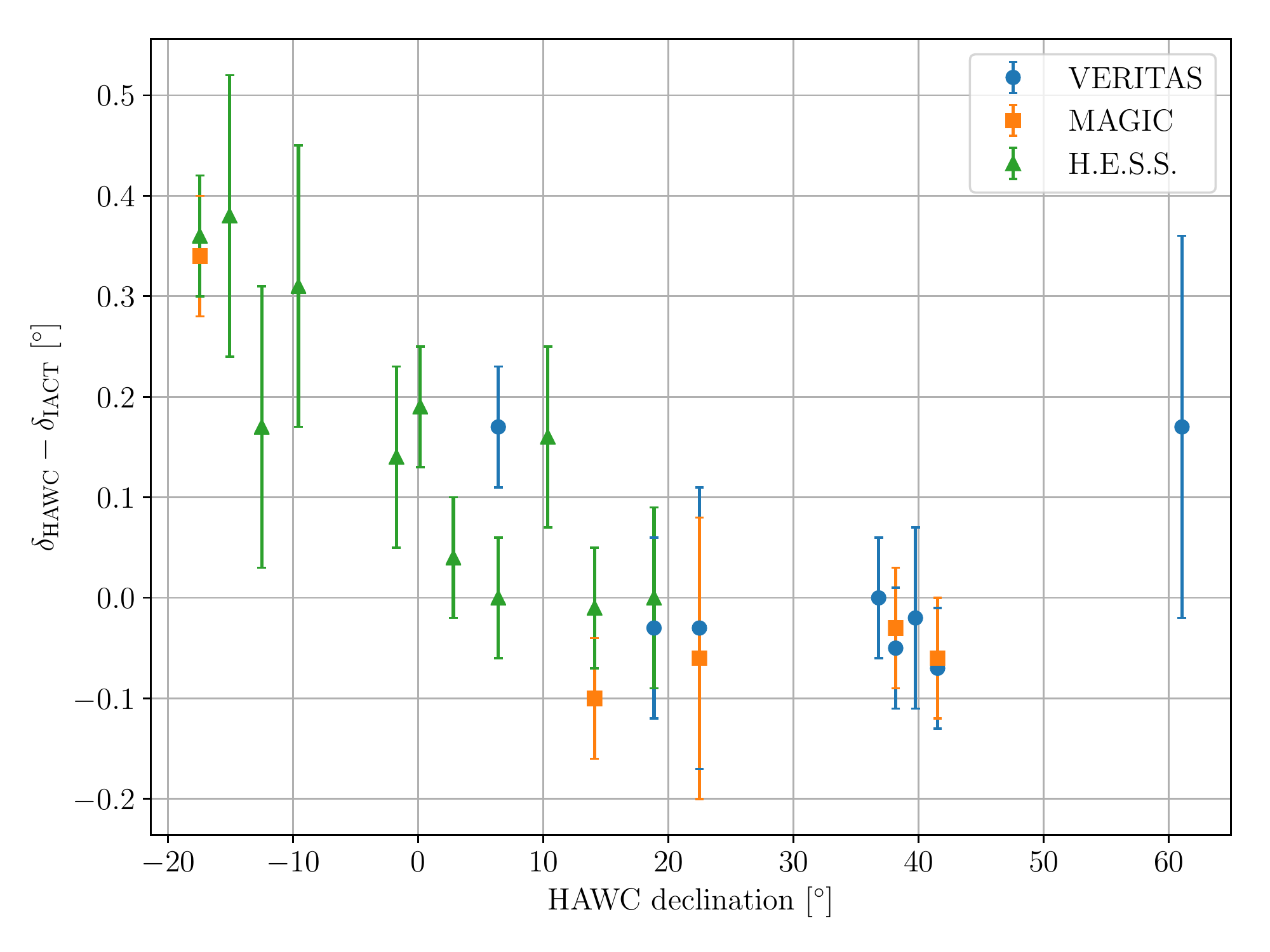}
\caption{Measured declination of HAWC sources relative to their TeV counterpart measurements from IACT experiments, MAGIC, H.E.S.S. and VERITAS. HAWC measurements agree with the source locations measured by IACTs within uncertainties for most of its declination range. See text for discussion on source declinations below \ang{-10}.} \label{fig:pointing}
\end{figure}

In Figure \ref{fig:pointing}, we compare the measured  declinations of  3HWC sources to the locations of their likely TeV counterparts as measured by other experiments. For this comparison, we consider relatively well-localised 3HWC sources that have a TeV association within \ang{1} detected by a different experiment. We do not include sources in regions of extended emission or multiple components such as \textbf{3HWC J2020+403}. It can be seen that for most of the declination range spanned by HAWC's sensitivity, the 3HWC positions agree with the literature values within statistical and systematic uncertainties. Below source declinations of about \ang{-10}, HAWC measures systematically higher values than the IACTs, between an offset of \ang{0.1} and \ang{0.4}.

The trend observed in Figure \ref{fig:pointing} could indicate a bias in HAWC's pointing at low declinations. However, all of HAWC's southern sources lie on the Galactic plane, in a region rich in sources and diffuse emission. HAWC's angular resolution is poor for low-declination sources compared to sources transiting overhead. Accordingly, Galactic diffuse emission or emission from nearby unresolved sources might affect the peak position detected by HAWC, especially for low-declination sources. IACTs tend to have better angular resolution and are thus affected less by large-scale emission or neighboring sources. The shift could also be an indication of an energy-dependent morphology of some sources \cite{Hona:2019ysf}. Future in-depth studies of some of these sources as well as an improved understanding of HAWC's pointing are needed to resolve this apparent discrepancy.

\subsection{Systematic Uncertainty of the Spectral Fits}
In Table \ref{tab:fluxes}, we report the best-fit fluxes and spectral indices of the 3HWC sources. These fits assume a power-law spectrum; we do not test other spectral models including a curvature or cutoff term. The reported spectral index should be interpreted as an average or effective spectral index across HAWC's energy range. For the two known extra-galactic sources, we also do not account for absorption by the extra-galactic background light. Additional studies are ongoing for sources that are detected with sufficient statistics to allow more sophisticated spectral models to be fit.

In Table \ref{tab:fluxes}, we report systematic uncertainties related to the modeling of the HAWC detector response individually for each source. More details about the sources of uncertainty considered here can be found in \cite{HighCrab}. In order to compute the uncertainties, we repeat spectral fits with certain properties of the detector model shifted up or down. We assign an additional uncertainty of 10\% to the flux normalization to account for effects, such as variations in the atmosphere, that are not considered otherwise. We add the resulting positive shifts to the spectral fit parameters in quadrature to obtain the total upward systematic uncertainty, and add the negative shifts in quadrature for the total negative downward systematic uncertainty.


There are other systematic issues affecting the spectral fits. Some fluxes may be overestimated due to ``leakage'' from nearby (detected or unresolved) sources or from the Galactic diffuse emission. These may also affect the spectral indices. In cases where the apparent extent of a source is larger than HAWC's angular resolution (\ang{0.1} at high energies), but the source is strong enough to be significantly detected already in the point-source search, we only report the spectrum assuming a point-source hypothesis. This leads to the flux normalization being underestimated and the spectral index to be biased towards softer spectra (as the angular resolution improves for high energies and thus more of the high-energy emission is ``lost''). Upcoming publications will provide better spectral fits for extended sources.


\section{Conclusion}\label{sec:conclude}
The HAWC observatory has been conducting the most sensitive, unbiased survey of the Northern sky at TeV energies for over five years. We have presented the third catalog of steady gamma-ray emitters detected by HAWC using 1523 days of data. The catalog consists of 65 sources, including two blazars. The most abundant source class among the potential counterpart of HAWC sources in the Galactic plane is pulsars (56). 

The 3HWC catalog provides many targets for multi-wavelength and multi-messenger follow-up studies that are crucial to several open problems in high-energy astrophysics. Detailed morphological and spectral studies of several sources are being conducted and will be the subject of future publications. A dedicated survey to constrain the emission from various extra-galactic objects of interest is under preparation. Future gamma-ray observatories such as CTA \citep{CTAConsortium:2018tzg} and SWGO \citep{Albert:2019afb} will be able to extend both the sensitivity and energy range of this survey.

\acknowledgments

We acknowledge the support from: the US National Science Foundation (NSF); the US Department of Energy Office of High-Energy Physics; the Laboratory Directed Research and Development (LDRD) program of Los Alamos National Laboratory; Consejo Nacional de Ciencia y Tecnolog\'ia (CONACyT), M\'exico, grants 271051, 232656, 260378, 179588, 254964, 258865, 243290, 132197, A1-S-46288, A1-S-22784, c\'atedras 873, 1563, 341, 323, Red HAWC, M\'exico; DGAPA-UNAM grants IG101320, IN111315, IN111716-3, IN111419, IA102019, IN112218; VIEP-BUAP; PIFI 2012, 2013, PROFOCIE 2014, 2015; the University of Wisconsin Alumni Research Foundation; the Institute of Geophysics, Planetary Physics, and Signatures at Los Alamos National Laboratory; Polish Science Centre grant, DEC-2017/27/B/ST9/02272; Coordinaci\'on de la Investigaci\'on Cient\'ifica de la Universidad Michoacana; Royal Society - Newton Advanced Fellowship 180385; Generalitat Valenciana, grant CIDEGENT/2018/034; Chulalongkorn University’s CUniverse (CUAASC) grant. Thanks to Scott Delay, Luciano D\'iaz and Eduardo Murrieta for technical support.




\bibliography{3hwc}{}
\bibliographystyle{aasjournal}



\end{document}